\documentclass[review]{elsarticle}

\usepackage{lineno,hyperref}
\usepackage{color}
\usepackage[utf8]{inputenc}
\usepackage{array}
\usepackage{graphicx}
\usepackage{subcaption}
\usepackage[normalem]{ulem}

\usepackage{amsfonts,amssymb,amsmath}

\modulolinenumbers[5]

\journal{Chemometrics and Intelligent Laboratory Systems}

\begin{document}

\begin{frontmatter}

\title{Robust and sparse estimation methods for 
high dimensional linear and logistic regression}

\author[sevinc]{Fatma Sevin\c{c} Kurnaz}
\cortext[Fatma Sevinc Kurnaz]{Corresponding author}
\ead{fskurnaz@yildiz.edu.tr, Tel: +902123837267}

\author[tuwien]{Irene Hoffmann}

\author[tuwien]{Peter Filzmoser}

\address[sevinc]{Department of Statistics, Yildiz Technical University, 34220, Istanbul, Turkey}

\address[tuwien]{Institute of Statistics and Mathematical Method in Economics, 
Vienna University of Technology, Wiedner Hauptstra\ss e 8-10, 1040, Vienna, Austria}

\begin{abstract}
Fully robust versions of the elastic net estimator are introduced for linear 
and logistic regression. The algorithms to compute the estimators 
are based on the idea of
repeatedly applying the non-robust classical estimators to data subsets only.
It is shown how outlier-free subsets can be identified efficiently, and how 
appropriate tuning parameters for the elastic net penalties can be selected.
A final reweighting step improves the efficiency of the estimators.
Simulation studies compare with non-robust and 
other competing robust estimators and reveal the superiority of the newly
proposed methods. This is also supported by a reasonable computation time and by
good performance in real data examples.
\end{abstract}

\begin{keyword}
Elastic net penalty, Least trimmed squares,
C-step algorithm, High dimensional data, Robustness, Sparse estimation
\end{keyword}

\end{frontmatter}

\linenumbers

\section{Introduction} 
\vskip-0.25cm 

Let us consider the linear regression model, which assumes the linear relationship 
between the predictors $\mathbf{X} \in \mathbb{R}^{n \times p}$ and the predictand 
$\mathbf{y} \in \mathbb{R}^{n \times 1}$,
\begin{equation}
\label{lr}
\mathbf{y}=\mathbf{X}\pmb{\beta}+\pmb{\varepsilon} ,
\end{equation}
where $\pmb{\beta}=(\beta_1,\ldots ,\beta_p)^T$ are the regression 
coefficients and $\pmb{\varepsilon}$ is the error 
term assumed to have standard normal distribution.
For simplicity we assume that $\mathbf{y}=(y_1,\ldots ,y_n)^T$ is centered to mean zero,
and the columns of $\mathbf{X}$ are mean-centered and scaled to variance one.
The ordinary least squares (OLS) regression estimator is the common choice
in situations where the number of observations, $n$, in the data set 
is greater than the number of predictor variables, $p$. 
However, in presence of multicollinearity among predictors, the OLS estimator 
becomes unreliable, and if $p$ exceeds $n$ it can not even be computed.
Several alternatives have been proposed in this case; here we focus on the 
class of shrinkage estimators which penalize the residual sum-of-squares.
The ridge estimator uses an $l_2$ penalty on the regression 
coefficients \cite{Hoerl70}, while the lasso estimator takes an 
$l_1$ penalty instead \cite{Tibshirani96}. Although this does no longer allow for
a closed form solution for the estimated regression coefficients, the lasso
estimator gets \textit{sparse}, which means that some of the regression coefficients
are shrunken to zero. This means that lasso acts like a variable selection 
method by returning a smaller subset of variables being relevant 
for the model. This is appropriate in particular for high dimensional low sample
size data sets  ($n \ll p$), arising from applications in chemometrics, biometrics, 
econometrics, social sciences and many other fields, where the data include
many uninformative variables which have no effect on the predictand or have very 
small contribution to the model.

There is also a limitation of the lasso estimator, since it is able to select only 
at most $n$ variables when $n<p$. If $n$ is very small, or if the number of 
informative variables (variables which are relevant for the model) is expected
to be greater than $n$, the model performance can become poor. As a way out,
the elastic net (\textit{enet}) estimator has been introduced \cite{Zou05}, which combines both
$l_1$ and $l_2$ penalties:
\begin{equation}
\label{elnet}
\hat{\pmb{\beta}}_{enet} =  \operatorname*{arg\,min}_{\pmb{\beta}}
\left\{\sum_{i=1}^n ( y_i-\mathbf{x}^T_i\pmb{\beta})^2 + \lambda P_{\alpha}(\pmb{\beta})
\right\}
\end{equation}
Here, $\mathbf{y}=(y_1,\ldots ,y_n)^T$,
the observations $\mathbf{x}_i^T$ form the rows of $\mathbf{X}$, and 
the penalty term $P_{\alpha}$ is defined as
\begin{equation} 
\label{penalty}
P_{\alpha}(\pmb{\beta})=(1-\alpha)\frac{1}{2} \lVert \pmb{\beta} \rVert_2^2 + \alpha \lVert \pmb{\beta} \rVert_1
=\sum_{j=1}^p \left[ (1-\alpha)\frac{1}{2} \beta_j^2 + \alpha \lvert \beta_j\rvert \right].
\end{equation}
The entire strength of the penalty is controlled 
by the tuning parameter $\lambda\geq 0$. The other tuning parameter $\alpha$ is the mixing 
proportion of the ridge and lasso penalties and takes value in $\left[0,1 \right]$.
The elastic net estimator is able to select variables like in lasso regression,
and shrink the coefficients according to ridge. For an overview of sparse methods, see \cite{Filzmoser12}. 

A further limitation of the previously mentioned estimators is their lack of robustness
against data outliers. 
In practice, the presence of outliers in data is quite common,
and thus robust statistical methods are frequently used, see, for example~\cite{Liang1,Liang2}.
In the linear regression setting, outliers may appear in the 
space of the predictand (so-called vertical outliers), or in the space of the 
predictor variables (leverage points) \cite{Maronna06}. The Least Trimmed Squares (LTS)
estimator has been among the first proposals of a regression estimator
being fully robust against both types of outliers \cite{RousseeuwL03}. 
It is defined as
\begin{equation}
\label{LTS}
\hat{\pmb{\beta}}_{LTS}= \operatorname*{arg\,min}_{\pmb{\beta}} \sum_{i=1}^hr_{(i)}^2(\pmb{\beta}) ,
\end{equation}
where the $r_{(i)}$ are the ordered absolute residuals 
$\lvert r_{(1)}\rvert \leq \lvert r_{(2)}\rvert \leq \dots \leq \lvert r_{(n)}\rvert$, 
and $r_{i}=y_i-\mathbf{x}_i^T\pmb{\beta}$ \cite{Rousseeuw84}. The number $h$ is
chosen between $\lfloor (n+p+1)/2\rfloor$ and $n$, where 
$\lfloor a \rfloor$ refers to the largest integer $\leq a$,
and it determines the robustness properties of
the estimator \cite{Rousseeuw84}. The LTS estimator also became popular
due to the proposal of a quick algorithm for its computation, the so-called
FAST-LTS algorithm \cite{Rousseeuw06}. The key feature of this algorithm is the 
``concentration step'' or C-step, which is an efficient way to arrive at outlier-free
data subsets where the OLS estimator can be applied. This only works for $n>p$, but
recently the sparse LTS regression estimator has been proposed for high dimensional
problems \cite{Alfons13}:
\begin{equation}
\label{spLTS}
\hat{\pmb{\beta}}_{sparseLTS} =  \operatorname*{arg\,min}_{\pmb{\beta}} 
\left\{\sum_{i =1}^h r_{(i)}^2(\pmb{\beta}) + h\lambda \lVert \pmb{\beta} \rVert_1
\right\}.
\end{equation}
This estimator adds an $l_1$ penalty to the objective function 
of the LTS estimator, and it can thus be seen as a robust counterpart of the 
lasso estimator. The sparse LTS estimator is robust to both vertical outliers and 
leverage points, and also a fast algorithm has been developed for its computation \cite{AlfonsR13}.

The contribution of this work is twofold: A new sparse and robust regression estimator
is proposed with combined $l_1$ and $l_2$ penalties. This robustified elastic net regression
estimator overcomes the limitations of lasso type estimators concerning the low number
of variables in the models, and concerning the instability of the estimator in case
of high multicollinearity among the predictors \cite{Tibshirani96}.
As a second contribution, a robust elastic net version of logistic regression is
introduced for problems where the response $\mathbf{y}$ is a binary variable, encoded
with $y_i\in \{0,1\}$ referring to the class memberships of two groups. 
The logistic regression model is $y_i=\pi_i+\varepsilon_i$, for $i=1,\ldots ,n$,
where $\pi_i$ denotes the conditional probability for the $i$th observation,
\begin{equation}
\pi_i=\mbox{Pr}(y_i=1|\mathbf{x}_i)=\frac{e^{\mathbf{x}^T_i\pmb{\beta}}}{1+e^{\mathbf{x}^T_i\pmb{\beta}}} ,
\label{eq:pi}
\end{equation}
and $\varepsilon_i$ is the error term assumed to have binomial distribution.
The most popular way to estimate the model parameters is the maximum likelihood (ML) 
estimator which is based on maximizing the log-likelihood
function or, equivalently, minimizing the negative log-likelihood function,
\begin{equation}
\label{MLlog}
\hat{\pmb{\beta}}_{ML} = \operatorname*{arg\,min}_{\pmb{\beta}}  
\sum_{i=1}^n d(\mathbf{x}^T_i\pmb{\beta},y_i),
\end{equation}
with the deviances
\begin{equation}
\label{deviances}
d(\mathbf{x}^T_i\pmb{\beta},y_i) = - y_i \log \pi_i - (1-y_i)\log (1-\pi_i) =
- y_i\mathbf{x}^T_i\pmb{\beta}+
\log\left(1+e^{\mathbf{x}^T_i\pmb{\beta}}\right) .
\end{equation}

The estimation of the model parameters with this method is not reliable when there is 
multicollinearity among the predictors and is not feasible when $p>n$. 
To solve these problems, Friedman et al. \cite{Friedman10} suggested to
minimize a penalized negative log-likelihood function, 
\begin{equation}
\label{enetlog}
\hat{\pmb{\beta}}_{enet} = \operatorname*{arg\,min}_{\pmb{\beta}}  
\left\{\sum_{i=1}^n d(\mathbf{x}^T_i\pmb{\beta},y_i) + 
n\lambda P_{\alpha}(\pmb{\beta}) \right\}.
\end{equation}
Here, $P_{\alpha}(\pmb{\beta})$ is the elastic net penalty as given in 
Equation (\ref{penalty}), and thus this estimator extends (\ref{elnet}) to the 
logistic regression setting.
Using the elastic net penalty also solves the non-existence problem of the 
estimator in case of non-overlapping groups \cite{Albert84,Friedman10,FriedmanR16}. 
Robustness can be achieved by trimming the penalized log-likelihood function, and 
using weights as proposed in the context of robust logistic regression  \cite{Croux03, Bianco96}.
These weights can also be applied in a reweighting step which increases the efficiency
of the robust elastic net logistic regression estimator.

The outline of this paper is as follows. 
In Section \ref{sec:enetlts}, we introduce the robust and sparse linear regression estimator 
and provide a detailed algorithm for its computation. 
Section \ref{sec:enetlogit} presents the robust elastic net logistic regression estimator.
Some important details which are different from the linear regression algorithm are 
mentioned here. Section \ref{sec:tuningpara} 
explains how the tuning parameters for the proposed estimators can be selected;
we prefer an approach based on cross-validation. 
Since LTS estimators possess a rather low statistical efficiency, a reweighting step
is introduced in Section~\ref{sec:rewight} to increase the efficiency.
The properties of 
the proposed estimators are investigated in simulation studies in Section \ref{sec:simulations}, 
and Section \ref{sec:applications}
shows the performance on real data examples.
Section \ref{sec:comptime} provides some insight into the computation time of the algorithms,
and the final Section \ref{sec:conclude} concludes.

\section{Robust and sparse linear regression with elastic net penalty}
\label{sec:enetlts}
\vskip-0.25cm

A robust and sparse elastic net estimator in linear regression can be defined with 
the objective function
\begin{equation}
\label{objectlinear}
Q(H,\pmb{\beta}) = \sum_{i \in H} ( y_i-\mathbf{x}^T_i\pmb{\beta})^2 + h\lambda P_{\alpha}(\pmb{\beta})
\end{equation} 
where $H \subseteq \{1,2,\dots,n\}$ with $\lvert H \rvert=h$, $\lambda \in \lbrack 0,\lambda_0 \rbrack$, and $P_{\alpha}$ indicates the elastic net penalty with $\alpha \in \lbrack 0,1 \rbrack$ as in Equation (\ref{penalty}). 
We call this estimator the \textit{enet-LTS} estimator, since it uses a trimmed sum of
squared residuals, like the sparse LTS estimator (\ref{spLTS}). The minimum of the 
objective function (\ref{objectlinear}) determines the optimal subset of size $h$,
\begin{equation}
\label{Hopt}
H_{opt} = \operatorname*{arg\,min}_{H \subseteq {1,2,\dots,n}:\lvert H \rvert=h} Q(H,\hat{\pmb{\beta}}_H) ,
\end{equation}
which is supposed to be outlier-free.
The coefficient estimates $\hat{\pmb{\beta}}_H$ depend 
on the subset $H$. 
The enet-LTS estimator is given for this subset $H_{opt}$ by 
\begin{equation}
\label{enetLTS}
\hat{\pmb{\beta}}_{enetLTS}=\operatorname*{arg\,min} Q(H_{opt},\pmb{\beta}). 
\end{equation}

It is not trivial to identify this optimal subset, and practically one has to use 
an algorithm to approximate the solution. This algorithm uses C-steps: Suppose that
the current $h$-subset in the $k$th iteration of the algorithm is denoted by $H_k$,
and the resulting estimator by $\hat{\pmb{\beta}}_{H_k}$.
Then the next subset $H_{k+1}$ is formed by the indexes of those observations which
correspond to the smallest $h$ squared residuals
\begin{equation}
\label{eq:CSTEPreg}
r^2_{k,i}=(y_i-\mathbf{x}^T_i\hat{\pmb{\beta}}_{H_k})^2, \quad \mbox{ for } i=1,\ldots ,n.
\end{equation}
If $\hat{\pmb{\beta}}_{H_{k+1}}$ denotes the estimator based on $H_{k+1}$, then
by construction of the $h$-subsets it follows immediately:
\begin{equation}
\label{eq:CSTEPobjf}
Q(H_{k+1},\hat{\pmb{\beta}}_{H_{k+1}})\leq Q(H_{k+1},\hat{\pmb{\beta}}_{H_{k}})
\leq Q(H_{k},\hat{\pmb{\beta}}_{H_{k}})
\end{equation}
This means that the C-steps decrease the objective function (\ref{objectlinear}) 
successively, and lead to a local optimum after convergence. The global optimum is
approximated by performing the C-steps with several initial subsets.
However, in order to keep the runtime of the algorithm low, it is crucial that the
initial subsets are chosen carefully.
As motivated in \cite{Alfons13}, for a certain combination of the penalty parameters 
$\alpha$ and $\lambda$, elemental subsets are created consisting of the indexes of 
three randomly selected observations. 
Using only three observations increases the possibility of having no outliers in the 
elemental subsets. Let us denote these elemental subsets by
\begin{equation}
\label{3obs}
H_{el}^s=\{j_1^s,j_2^s,j_3^s\} ,
\end{equation}
where $s \in \{1,2,\dots,500 \}$. The resulting estimators based on the three
observations are denoted by $\hat{\pmb{\beta}}_{H_{el}^s}$.
Now the squared residuals $(y_i-\mathbf{x}_i \hat{\pmb{\beta}}_{H_{el}^s})^2$
can be computed for all observations $i=1,\ldots ,n$, and two C-steps are 
carried out, starting with the $h$-subset defined by the indexes of the smallest 
squared residuals.
Then only those $10$ $h$-subsets with the smallest values of the objective 
function (\ref{objectlinear}) are kept as candidates. 
With these candidate subsets, the C-steps are performed until convergence
(no further decrease), and 
the best subset is defined as that one with the smallest value of the objective function.
This \textit{best subset} also defines the estimator for this particular combination of
$\alpha$ and $\lambda$.

Basically, one can apply this procedure now for a grid of values in the
interval $\alpha \in \lbrack 0,1 \rbrack$ and $\lambda \in \lbrack 0,\lambda_0 \rbrack$.
Practically, this may still be quite time consuming, and therefore, for a new 
parameter combination, the best subset of the neighboring
grid value of $\alpha$ and/or $\lambda$, is taken, and the C-steps are started from
this best subset until convergence. 
This technique, called \textit{warm starts}, is repeated for each combination over 
the grid of $\alpha$ and $\lambda$
values, and thus the start based on the elemental subsets is carried out only once.

The choice of the optimal tuning parameters $\alpha_{opt}$ 
and $\lambda_{opt}$ is detailed in Section \ref{sec:tuningpara}.  
The subset corresponding to the optimal tuning parameters is the 
optimal subset of size $h$.
The enet-LTS estimator is then calculated on the optimal subset with $\alpha_{opt}$ 
and $\lambda_{opt}$.

\section{Robust and sparse logistic regression with elastic net penalty}
\label{sec:enetlogit}
\vskip-0.25cm

Based on the definition (\ref{enetlog}) of the elastic net logistic regression estimator,
it is straightforward to define the objective function of its robust counterpart
based on trimming,
\begin{equation}
\label{objlog}
Q(H,\pmb{\beta}) = \sum_{i \in H}
d(\mathbf{x}^T_i\pmb{\beta},y_i) + h\lambda P_{\alpha}(\pmb{\beta}) ,
\end{equation}
where again $H \subseteq \{1,2,\dots,n\}$ with $\lvert H \rvert=h$, and $P_{\alpha}$ is 
the elastic net penalty as defined in Equation (\ref{penalty}). 
As outlined in the last Section \ref{sec:enetlts}, the task is to find the optimal subset
which minimizes the objective function and defines the robust sparse elastic net
estimator for logistic regression. It turns out that the algorithm explained previously
in the linear regression setting can be successfully used to find the approximative
solution. In the following we will explain the modifications that need to be carried out.
\begin{description}
\item[C-steps:] In the linear regression case, the C-steps were based on the squared
residuals (\ref{eq:CSTEPreg}).
Now the $h$-subsets are determined according to the 
indexes of those observations with the smallest values 
of the deviances $d(\mathbf{x}^T_i\hat{\pmb{\beta}}_{H_{k}},y_i)$.
However, here it needs to be made sure that the original group sizes are in the 
same proportion. Denote $n_0$ and $n_1$ the number of observations in both groups,
with $n_0+n_1=n$. Then $h_0=\lfloor (n_0+1) h/n\rfloor$ and $h_1=h-h_0$ define the 
group sizes in each $h$-subset. A new $h$-subset is created with the 
$h_0$ indexes of the smallest deviances $d(\mathbf{x}^T_i\hat{\pmb{\beta}}_{H_{k}},y_i=0)$
and with the $h_1$ indexes of the smallest deviances $d(\mathbf{x}^T_i\hat{\pmb{\beta}}_{H_{k}},y_i=1)$.

\item[Elemental subsets:] In the linear regression case, the elemental subsets
consisted of the indexes of three randomly selected observations, see (\ref{3obs}).
Now four observations are randomly selected to form the elemental subsets, 
two from each group. This allows to compute the estimator, and the two C-steps 
are based on the $h$ smallest values of the deviances.
As before, this is carried out for 500 elemental subsets, and only the ``best'' 10
$h$-subsets are kept.
Here, ``best'' refers to an evaluation that is borrowed from a robustified deviance
measure proposed in Croux and Haesbroeck \cite{Croux03} in the context of robust logistic
regression (but not in high dimension). 
These authors replace the deviance function (\ref{deviances}) used in 
(\ref{MLlog}) by a function $\varphi_{BY}$ to define the so-called Bianco Yohai (BY)
estimator
\begin{equation}
\hat{\pmb{\beta}}_{BY} = \operatorname*{arg\,min}_{\pmb{\beta}} \sum_{i=1}^n \varphi(\mathbf{x}^T_i\pmb{\beta};y_i) ,
\label{BY}
\end{equation}
a highly robust logistic regression estimator, see also \cite{Bianco96}.
The form of the function $\varphi_{BY}$ is shown in Figure~\ref{fig:phifunc}, see
\cite{Croux03} for details.

We use this function as follows: Positive scores $\mathbf{x}^T_i\hat{\pmb{\beta}}$ of group 1, 
i.e. $y_i=1$,
refer to correct classification and receive the highest values for $\varphi_{BY}$,
while negative scores refer to misclassification, with small or zero $\varphi_{BY}$ 
values. For the scores of group 0 we have the reverse behavior, see Figure~\ref{fig:phifunc}.
When evaluating an $h$-subset, the sum over the $h$ values of
$\varphi_{BY}(\mathbf{x}^T_i\hat{\pmb{\beta}}_H)$ for $i\in H$
is computed, and this sum should be as large as possible. This means that we aim at
identifying an $h$-subset where the groups are separated as much as possible.
Points on the wrong side have almost no contribution, but also the contribution
of outliers on the correct side is bounded. In this way, outliers will not dominate the
sum.

With the best 10 $h$-subsets we continue the C-steps until convergence. 
Finally, the subset with the largest sum 
$\varphi_{BY}(\mathbf{x}^T_i\hat{\pmb{\beta}}_H)$ over all $i\in H$
forms the best index set. 

\begin{figure}[htp]
\begin{center}
\includegraphics[width=0.7\textwidth]{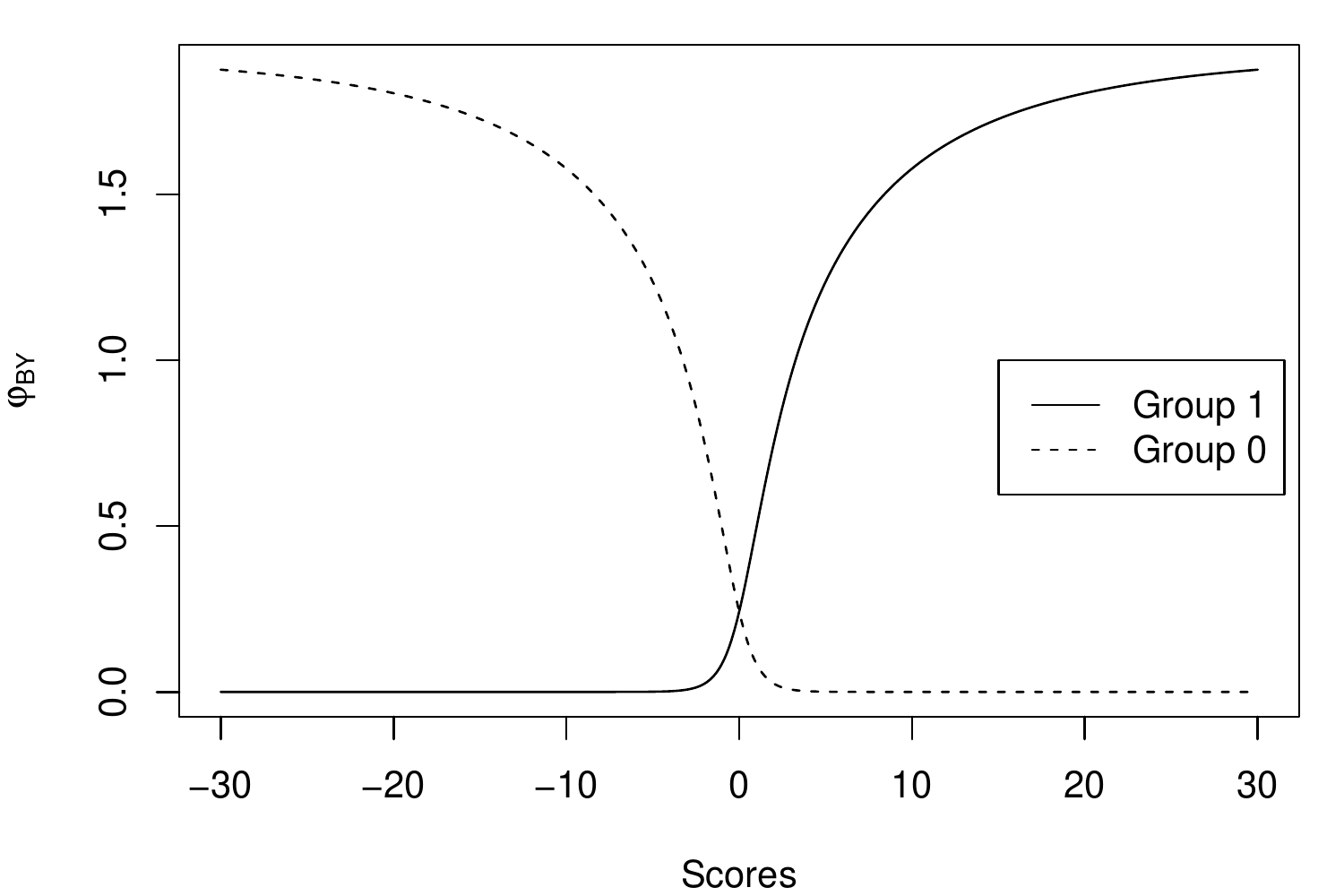}
\caption{\label{fig:phifunc}Function $\varphi_{BY}$ used for evaluating an $h$-subset, based
on the scores $\mathbf{x}^T_i\hat{\pmb{\beta}}$ for the two groups.}
\end{center}
\end{figure}

\end{description}
The selection of the optimal parameters $\alpha_{opt}$ 
and $\lambda_{opt}$ is discussed in Section \ref{sec:tuningpara}.  
The subset corresponding to these optimal tuning parameters is defined as the 
optimal subset of size $h$.
The enet-LTS logistic regression estimator is then calculated on the optimal subset
with $\alpha_{opt}$ and $\lambda_{opt}$.

\medskip
Note that at the beginning of the algorithm for linear regression, 
the predictand is centered, and the predictor variables are centered robustly by the median 
and scaled by the MAD. Within the C-steps of the algorithm, we additionally mean-center the 
response variable and scale the predictors by their arithmetic means and standard 
deviations, calculated on each current subset, see also \cite{Alfons13}. 
The same procedure is applied for logistic regression, except for centering 
the predictand.
In the end, the coefficients are back-transformed to the original scale.

\section{Selection of the tuning parameters}
\label{sec:tuningpara}
\vskip-0.25cm

Sections \ref{sec:enetlts} and \ref{sec:enetlogit} outlined the algorithms to arrive at a best 
subset for robust elastic net linear and logistic regression, for each combination
of the tuning parameters $\alpha \in [0,1]$ and $\lambda \in [0,\lambda_0]$.
In this section we define the strategy to select the optimal combination
$\alpha_{opt}$ and $\lambda_{opt}$, leading to the optimal subset.
For this purpose we are using $k$-fold cross-validation (CV) on those 
best subsets of size $h$, with $k=5$.
In more detail, for $k$-fold CV, the data are randomly split into $k$ blocks of
approximately equal size. In case of logistic regression, each block needs to consist 
of observations from both classes with approximately the same class proportions as in 
the complete data set. Each block is left out once, the model is fitted to the ``training data''
contained in the $k-1$ blocks,
using a fixed parameter combination for $\alpha$ and $\lambda$, and it is applied to
the left-out block with the ``test data''. In this way, $h$ fitted values are obtained
from $k$ models, 
and they are compared 
to the corresponding original response by using the following evaluation criteria:
\begin{itemize}
\item For linear regression we take the root mean squared prediction error 
(RMSPE)
\begin{equation}
\label{eq:cvrmspe}
\mathrm{RMSPE}(\alpha,\lambda) =\sqrt{\frac{1}{h}\sum_{i=1}^{h} r_i^2
(\hat{\pmb{\beta}}_{\alpha,\lambda})}
\end{equation}
where 
$r_{i}=y_i-\mathbf{x}^T_i\hat{\pmb{\beta}}_{\alpha,\lambda}$ presents the test set 
residuals from the models estimated on the training sets with a specific 
$\alpha$ and $\lambda$ (for simplicity we omitted here the index $k$ denoting the 
models where the $k$-th block was left out and the corresponding test data from this block). 
 
\item
For logistic regression we use the mean of the negative log-likelihoods 
or deviances (MNLL)
\begin{equation}
\label{eq:cvlog}
\mathrm{MNLL}(\alpha,\lambda) = \frac{1}{h}\sum_{i=1}^{h} d_{i}(\hat{\pmb{\beta}}_{\alpha,\lambda}),
\end{equation}
where 
$d_{i}=d(\mathbf{x}^T_i\hat{\pmb{\beta}}_{\alpha,\lambda},y_i)$
presents the test set deviances from the models estimated on the training sets with 
a specific $\alpha$ and $\lambda$.
\end{itemize}
Note that the evaluation criteria given by (\ref{eq:cvrmspe}) and (\ref{eq:cvlog}) are robust
against outliers, because they are based on the best subsets  of size $h$, which are supposed to
be outlier free.

In order to obtain more stable results, we repeat the $k$-fold CV five times and take the 
average of the corresponding evaluation measure.
Finally, the optimal parameters $\alpha_{opt}$ and $\lambda_{opt}$ are defined as that 
couple for which the evaluation criterion gives the minimal value. The corresponding 
best subset is determined as the optimal subset.

Note that the optimal couple $\alpha_{opt}$ and $\lambda_{opt}$ is searched on a grid 
of values $\alpha \in [0,1]$ and $\lambda \in [0,\lambda_0]$. 
In our experiments we used 41 equally spaced values for $\alpha$, and $\lambda$ was varied in
steps of size $0.025\lambda_0$. For determining $\lambda_0$ in the linear regression case 
we used the same approach as in Alfons et al.~\cite{Alfons13} which is based on the 
Pearson correlation between $y$ and the $j$th predictor variable $x_j$ on winsorized data.
For logistic regression we replaced the Pearson correlation by a robustified point-biserial correlation: denote by $n_0$ and $n_1$ the group sizes of the two groups, and by
$m_j^0$ and $m_j^1$ the medians of the $j$th predictor variable for the data from the 
two groups, respectively. Then the robustified point-biserial correlation between $y$ and 
$x_j$ is defined as
$$
r_{pb}(y,x_j)=\frac{m_j^1-m_j^0}{\mbox{MAD}(x_j)}\cdot \sqrt{\frac{n_0n_1}{n(n-1)}} ,
$$
where $\mbox{MAD}(x_j)$ is the MAD of $x_j$, and $n=n_0+n_1$.

\section{Reweighting step}
\label{sec:rewight}
\vskip-0.25cm

The LTS estimator has a low efficiency, and thus it is common to use a reweighting step
\cite{RousseeuwL03}. This idea is also used for the estimators introduced here.
Generally, in a reweighting step the outliers according to the current model are identified
and downweighted.
For the linear regression model we will use the same reweighting 
scheme as proposed in Alfons et al. \cite{Alfons13}, which is based on standardized residuals.
In case of logistic regression we compute the Pearson residuals which are approximately standard 
normally distributed and given by
\begin{equation}
r_i^s=\frac{y_i-\pi_i}{\pi_i\left(1-\pi_i\right)} ,
\label{pearson}
\end{equation}
with $\pi_i$ the conditional probabilities from (\ref{eq:pi}). 

For simplicity, denote the standardized residuals from the linear regression case also by $r_i^s$.
Then the weights are defined by 
\begin{equation}
w_i=\begin{cases}
1, & \mbox{ if  } \lvert r_i^s \rvert \leq \Phi^{-1}(1-\delta) \\
0, & \mbox{ if } \lvert r_i^s \rvert > \Phi^{-1}(1-\delta)
\end{cases} 
 \quad i=1,2,\dots,n,
\end{equation}
where $\delta=0.0125$, such that $2.5\%$ of the observations are flagged as outliers in 
the normal model. 
The reweighted enet-LTS estimator is defined as 
\begin{equation}
\label{eq:rewest}
\hat{\pmb{\beta}}_{reweighted} = \operatorname*{arg\,min}_{\pmb{\beta}}
\left\{\sum_{i=1}^n w_i f(\mathbf{x}_i;y_i) + \lambda_{upd} n_w P_{\alpha_{opt}}(\pmb{\beta})
\right\},
\end{equation}
where $w_i$, $i=1,\dots,n$ stands for the vector of binary weights (according to the current model), 
$n_w=\sum_{i=1}^n w_i$, and $f$ corresponds to squared residuals for linear regression or 
to the deviances in case of logistic regression.
Since $h \leq n_w$, and because the optimal parameters $\alpha_{opt}$ and $\lambda_{opt}$
have been derived with $h$ observations, the penalty can act (slightly) differently in
(\ref{eq:rewest}) than for the raw estimator. For this reason, the parameter $\lambda_{opt}$
has to be updated, while the $\alpha_{opt}$ regulating the tradeoff between the $l_1$ and
$l_2$ penalty is kept the same. The updated parameter $\lambda_{upd}$ is determined
by $5$-fold CV, with the simplification that
$\alpha_{opt}$ is already fixed.

\section{Simulation studies}
\label{sec:simulations}
\vskip-0.25cm 

In this section, the performance of the new estimators is compared with different sparse 
estimators in different scenarios. 
We consider both the raw and the reweighted versions of the enet-LTS estimators, and
therefore aim to show how the reweighting step improves the methods. 
The raw and reweighted enet-LTS estimators are compared with their classical, non-robust 
counterparts, the linear and logistic regression estimators with elastic net penalty \cite{Friedman10}.
In case of linear regression we also compare with the reweighted
sparse LTS estimator of \cite{Alfons13}. All robust estimators are calculated taking the 
subset size $h=\lfloor (n+1)\cdot 0.75\rfloor$ such that their performances are directly
comparable.

For each replication, we choose the optimal tuning parameters $\alpha_{opt}$ and $\lambda_{opt}$ 
over the grids $\alpha$ and $\lambda$ with 5-times repeated
$5$-fold CV as described in Section \ref{sec:tuningpara}. 
To select the tuning parameters for the classical estimators with elastic net penalty, 
we first draw the same 
grid for $\alpha$, namely $\alpha \in [0,1]$, with 41 equally spaced grid points. 
Then we use $5$-fold CV as provided by the R package \textit{glmnet}, 
which automatically checks the model quality 
for a sequence of values for $\lambda$,
taking the mean squared error
as an evaluation criterion. 
Finally, the tuning parameters corresponding to the smallest value of the 
minimum cross-validated error are determined as the optimal tuning parameters.
In order to be coherent with our evaluation, the tuning parameters for the sparse LTS estimator 
are determined in the same way as for the enet-LTS estimator.
All simulations are carried out in R \cite{R}.

Note that we simulated the data sets with intercept. As described at 
the end of Section \ref{sec:enetlogit}, the data are centered and scaled at the beginning of 
the algorithm and only in the final step the coefficients are back-transformed to the 
original scale, where also the estimate of the intercept is computed.


\textbf{\textsl{Sampling schemes for linear regression}:}
Let us consider two different scenarios  by means of generating a ``low dimensional'' data set 
with $n=150$ and $p=60$ and a ``high dimensional'' data set with 
$n=50$ and $p=100$. 
We generate a data matrix where the variables are forming correlated blocks,
$\mathbf{X}=(\mathbf{X}_{a_1},\mathbf{X}_{a_2},\mathbf{X}_b)$, where $\mathbf{X}_{a_1}$, $\mathbf{X}_{a_2}$ and $\mathbf{X}_b$ have the dimensions 
$n \times p_{a_1}$,$n \times p_{a_2}$ and $n \times p_b$, with $p=p_{a_1}+p_{a_2}+p_b$.
Such a block structure can be assumed in many application, and it mimics different 
underlying hidden processes. The observations of the blocks are generated independently 
from each other, from a multivariate normal distribution
$\mathcal{N}_{p_{a_1}}(\mathbf{0},\mathbf{\Sigma}_{a_1})$ with 
$\mathbf{\Sigma}_{a_1}=\rho_{a_1}^{\lvert j-k \rvert}$, $1 \leq j$, 
$k \leq p_{a_1}$, $\mathcal{N}_{p_{a_2}}(\mathbf{0},\mathbf{\Sigma}_{a_2})$ with 
$\mathbf{\Sigma}_{a_2}=\rho_{a_2}^{\lvert j-k \rvert}$, $1 \leq j$, 
$k \leq p_{a_2}$, and $\mathcal{N}_{p_b}(\mathbf{0},\mathbf{\Sigma}_b)$ 
with $\mathbf{\Sigma}_b=\rho_b^{\lvert j-k \rvert}$, $1 \leq j$, $k \leq p_b$, respectively.
While the first two blocks belong to the informative variables with sizes of $p_{a_1}=0.05p$ and $p_{a_2}=0.05p$, the third block represents uninformative variables with $p_b=0.9p$. 
Furthermore, we take $\rho_{a_1}=\rho_{a_2}=0.9$ to allow for a high correlation among 
the informative variables, and $\rho_b=0.2$ to have low correlation 
among the uninformative variables.

To create sparsity, the true parameter vector $\pmb{\beta}$ consists of zeros for the last
90\% of the entries referring to the uninformative variables, while the first 10\% of the 
entries are assigned to one.
The response variable is calculated by
\begin{equation}
y_i=1+\mathbf{x}_i^T\pmb{\beta}+\varepsilon_i ,
\label{eq:predictand_lin}
\end{equation}
where the error term $\varepsilon_i$ is distributed according to a standard 
normal distribution $\mathcal{N}(0,1)$, for $i=1,\dots,n$.

This is the design for the simulations with clean data.
For the simulation scenarios with outliers we replace the first $10\%$ of the observations 
of the block of informative variables by values coming from independent normal 
distributions $\mathcal{N}(20,1)$ for each variable.
Further, the error terms for these $10\%$ outliers are replaced by values from 
$\mathcal{N}(20\hat{\sigma}_y,1)$ instead of $\mathcal{N}(0,1)$, where 
$\hat{\sigma}_y$ represents 
the estimated standard deviation of the clean predictand vector.
In this way, the contaminated data consist of both vertical outliers and leverage points.

\textbf{\textsl{Sampling schemes for logistic regression}:}
We also consider two different scenarios for logistic regression, a ``low dimensional'' 
data set 
with $n=150$ and $p=50$ 
and a ``high dimensional'' data set with 
$n=50$ and $p=100$. 
The data 
matrix is $\mathbf{X}=(\mathbf{X}_a,\mathbf{X}_b)$, where $\mathbf{X}_a$ has the dimension 
$n \times p_a$ and $\mathbf{X}_b$ is of dimension $n \times p_b$, with $p=p_a+p_b$.
The data matrices are generated independently 
from $\mathcal{N}_{p_a}(\mathbf{0},\mathbf{\Sigma}_a)$ with 
$\mathbf{\Sigma}_a=\rho_a^{\lvert j-k \rvert}$, $1 \leq j$, 
$k \leq p_a$, and $\mathcal{N}_{p_b}(\mathbf{0},\mathbf{\Sigma}_b)$ 
with $\mathbf{\Sigma}_b=\rho_b^{\lvert j-k \rvert}$, $1 \leq j$, $k \leq p_b$, respectively. 
While the first block consists of the informative variables with $p_a=0.1p$, the second block represents uninformative variables with $p_b=0.9p$. 
We take $\rho_a=0.9$ for a high correlation among 
the informative variables, and $\rho_b=0.5$ for moderate correlation 
among the uninformative variables.

The coefficient vector $\pmb{\beta}$ consists of ones for the first 10\% 
of the entries, and zeros for the remaining uninformative block.
The elements of the error term $\varepsilon_i$ are generated independently from
$\mathcal{N}(0,1)$. 
The grouping variable is then generated according to the model
\begin{equation}
\label{eq:predictand}
y_i=\begin{cases}
0, & \mbox{ if } 1+\mathbf{x}_i^T\pmb{\beta}+\varepsilon_i \leq 0\\
1, & \mbox{ if } 1+\mathbf{x}_i^T\pmb{\beta}+\varepsilon_i > 0
\end{cases}
\quad i=1,2,\dots,n.
\end{equation}
With this setting, both groups are of approximately the same size.

Contamination is introduced by adding outliers only to the informative variables.
Denote $n_0$ the number of observations in class 0.
Then the first $\lfloor 0.1 n_0 \rfloor$ observations of group 0 are replaced by
values generated from $\mathcal{N}(20,1)$.
In order to create ``vertical'' outliers in addition to leverage points, 
we assign those first $0.1 n_0$ observations 
of class 0 a wrong class membership.

\textbf{\textsl{Performance measures}:}
For the evaluation of the different estimators, training and test data sets are generated 
according to the explained sampling schemes. The models are fit to the training data
and evaluated on the test data. The test data are always generated without outliers.

As performance measures we use the root mean squared prediction error (RMSPE) for linear regression,
\begin{equation}
\label{eq:rmspe}
\mathrm{RMSPE}(\hat{\pmb{\beta}})=\sqrt{\frac{1}{n}\sum_{i=1}^n\left(y_i-\hat{\beta}_0-\mathbf{x}_i^T\hat{\pmb{\beta}} \right)^2} ,
\end{equation}
and the mean of the negative log-likelihoods or deviances (MNLL) for logistic regression, 
\begin{equation}
\label{eq:nll}
\mathrm{MNLL}(\hat{\pmb{\beta}}) = \frac{1}{n}\sum_{i=1}^{n} 
d(\hat{\beta}_0+\mathbf{x}^T_i\hat{\pmb{\beta}},y_i) ,
\end{equation}
where $y_i$ and $\mathbf{x}_i$, $i=1,\dots,n$, indicate the observations of the test data set, 
$\hat{\pmb{\beta}}$ denotes the coefficient vector and
 $\hat{\beta}_0$ stands for the estimated intercept term obtained 
from the training data set.
In logistic regression we also calculate the 
misclassification rate (MCR), defined as
\begin{equation}
\mathrm{MCR}=\frac{m}{n}
\label{eq:mcr}
\end{equation}
where $m$ is the number of misclassified observations from the test data after fitting the 
model on the training data. 
Further, we consider the precision of the coefficient estimate 
as a quality criterion, defined by
\begin{equation}
\label{eq:bias}
\mathrm{PRECISION}(\hat{\pmb{\beta}})=\sqrt{\sum_{i=0}^{p}\left(\beta_i-\hat{\beta}_i 
\right)^2},
\end{equation}
In order to compare the sparsity of the coefficient estimators, we evaluate the 
False Positive Rate (FPR) and the False Negative Rate (FNR), defined as
\begin{equation}
\label{eq:fpr}
\mathrm{FPR}(\hat{\pmb{\beta}})=\frac{\lvert \{ j=0,\dots,p:\hat{\beta}_j \neq 0 \wedge \beta_j=0 \} \rvert}{\lvert \{ j=0,\dots,p:\beta_j=0\}  \rvert},
\end{equation}
\begin{equation}
\label{eq:fnr}
\mathrm{FNR}(\hat{\pmb{\beta}})=\frac{\lvert \{ j=0,\dots,p:\hat{\beta}_j=0 \wedge \beta_j \neq 0 \} \rvert}{\lvert \{ j=0,\dots,p:\beta_j \neq 0\}  \rvert}.
\end{equation}
The FPR is the proportion of non-informative variables that are incorrectly included
in the model. On the other hand, the FNR is the proportion of informative variables that
are incorrectly excluded from the model.
A high FNR usually has a bad effect on the prediction performance since it 
inflates the variance of the estimator.
 
These evaluation measures are calculated for the generated data in each of 
100 simulation replications separately, 
and then summarized by boxplots. 
The smaller the value for these criteria, the better the performance of the method.

\begin{figure}[htbp]
\includegraphics[width=0.5\textwidth]{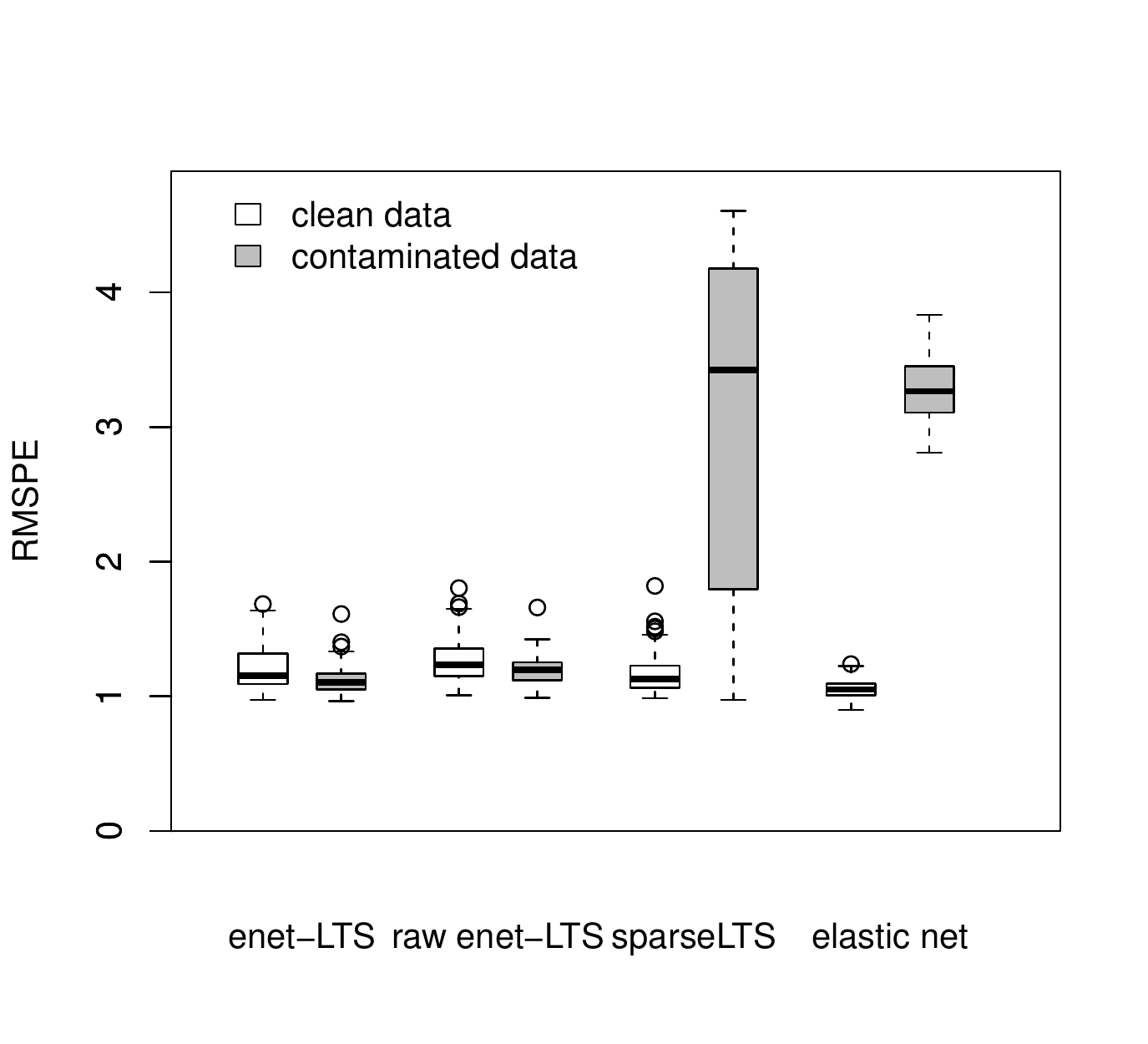}
\hfill
\includegraphics[width=0.5\textwidth]{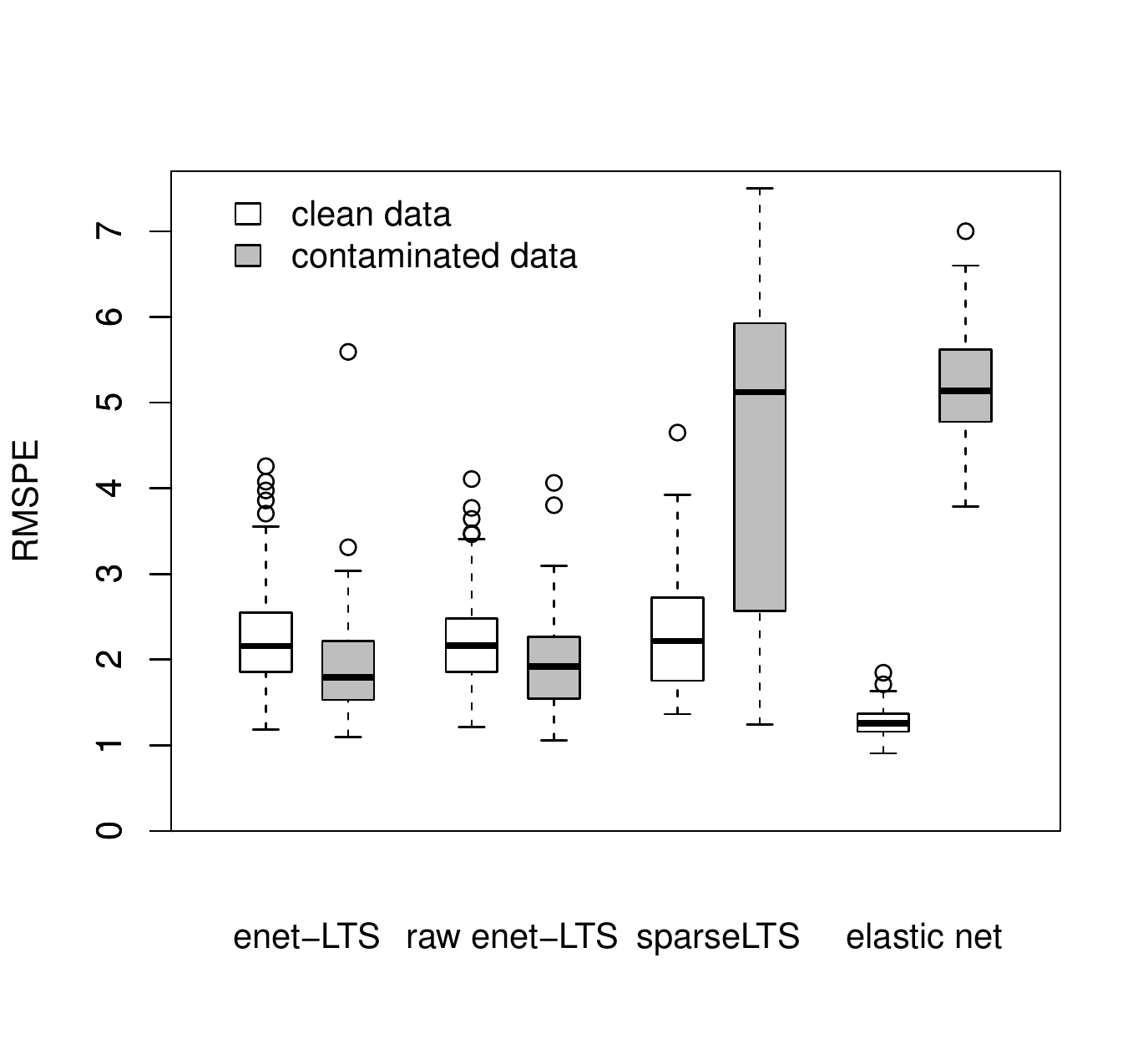}
\caption{Root mean squared prediction error (RMSPE) for linear regression.
Left: low dimensional data set ($n=150$ and $p=60$); right: high dimensional data set ($n=50$ and $p=100$).}
\label{fig:rmspe_lin}
\end{figure}

\begin{figure}[htbp]
\includegraphics[width=0.5\textwidth]{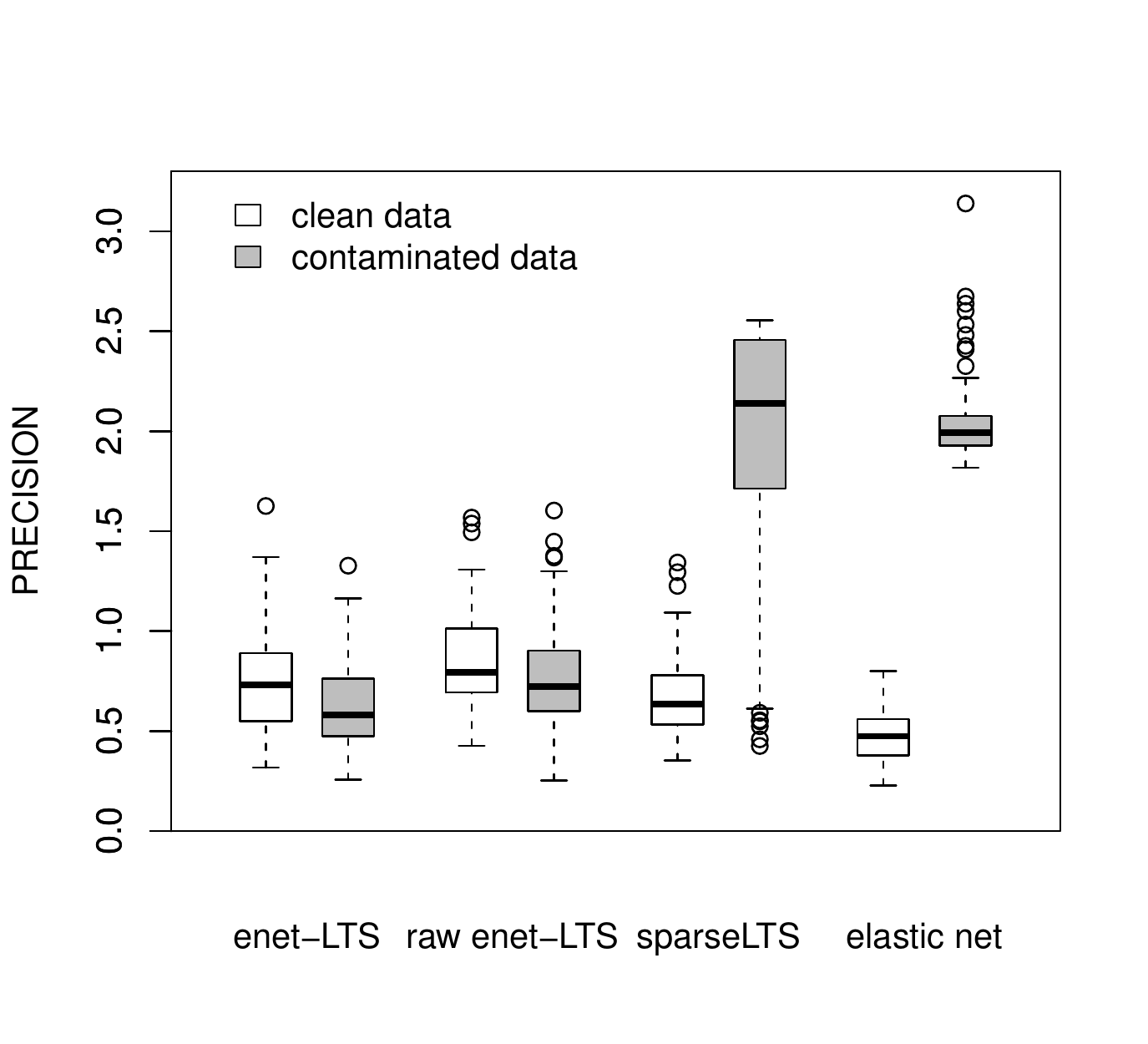}
\hfill
\includegraphics[width=0.5\textwidth]{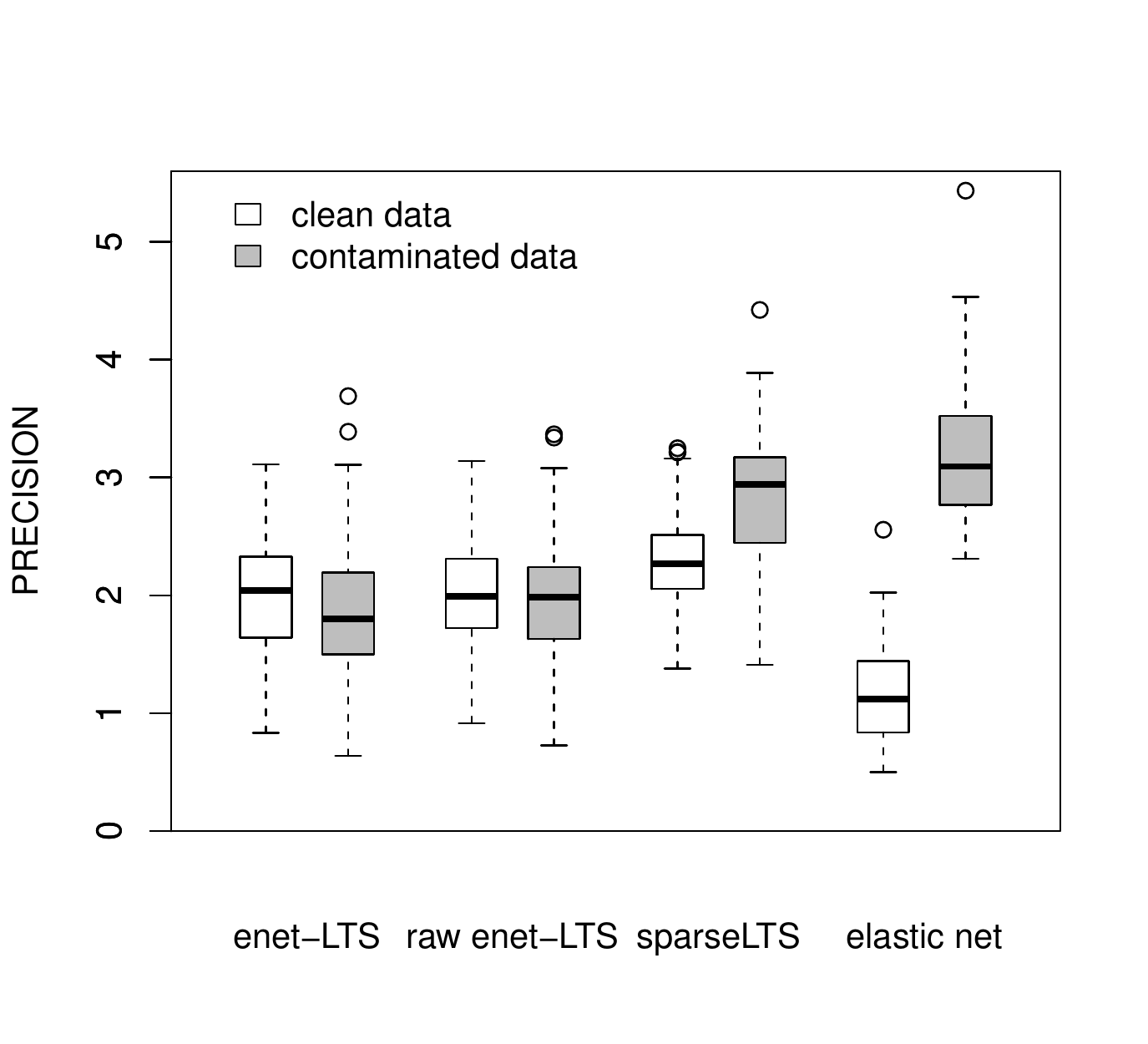}
\caption{Precision of the estimators (PRECISION) for linear regression.
Left: low dimensional data set ($n=150$ and $p=60$); right: high dimensional data set ($n=50$ and $p=100$).}
\label{fig:bias_lin}
\end{figure}

\begin{figure}[htbp]
\includegraphics[width=0.5\textwidth]{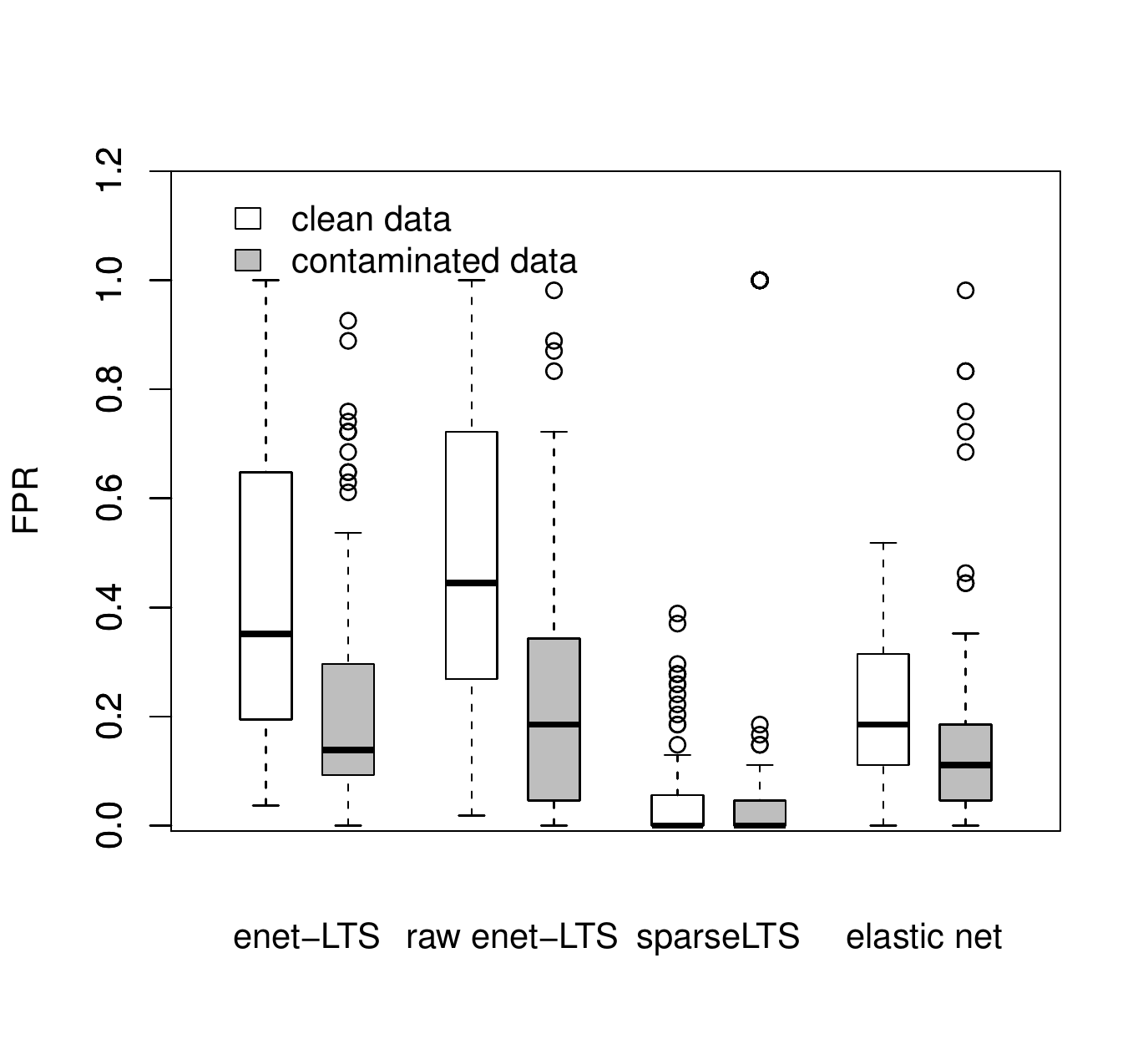}
\hfill
\includegraphics[width=0.5\textwidth]{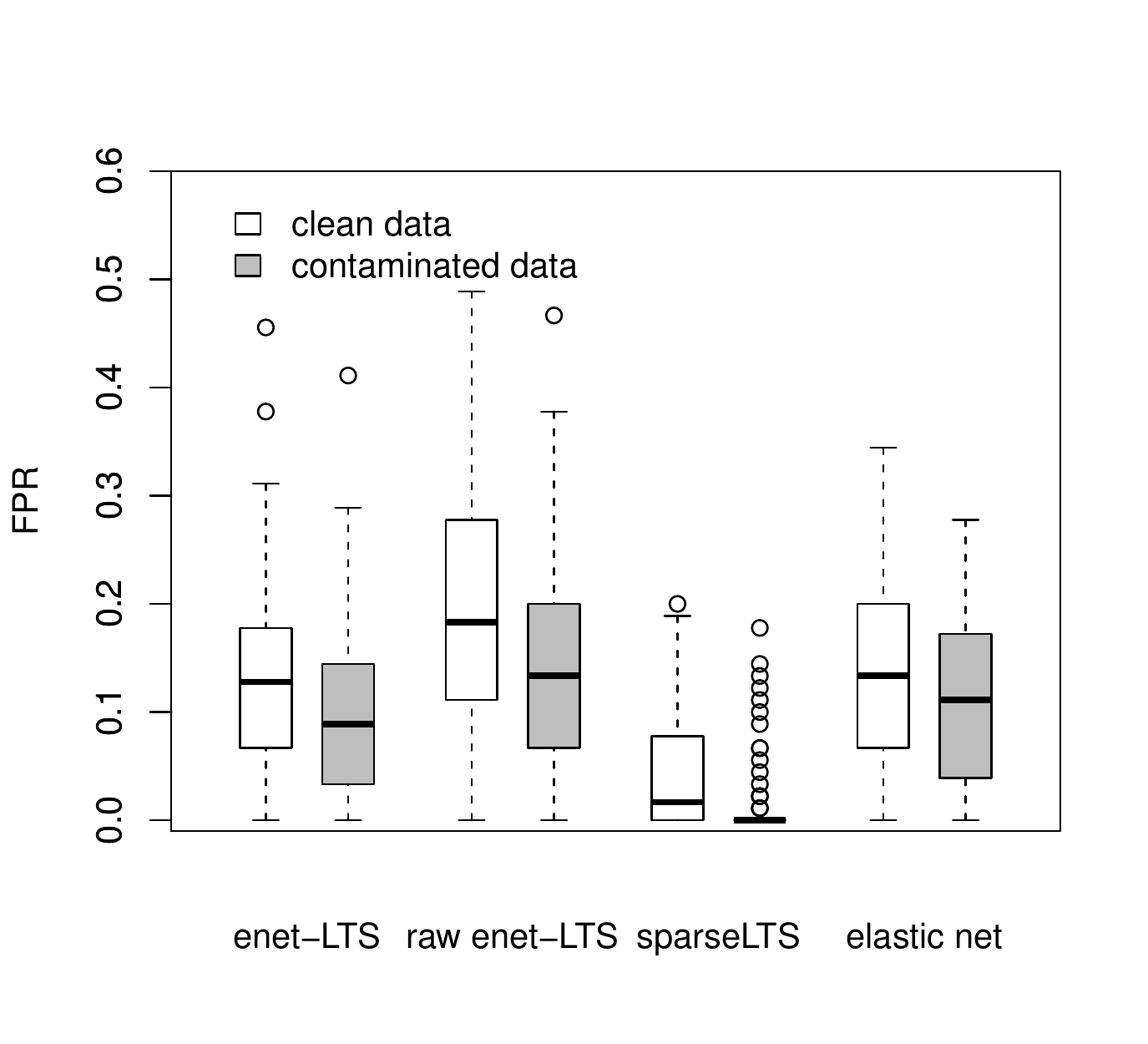}
\caption{False positive rate (FPR) for linear regression.
Left: low dimensional data set ($n=150$ and $p=60$); right: high dimensional data set ($n=50$ and $p=100$).}
\label{fig:fpr_lin}
\end{figure}

\begin{figure}[htbp]
\includegraphics[width=0.5\textwidth]{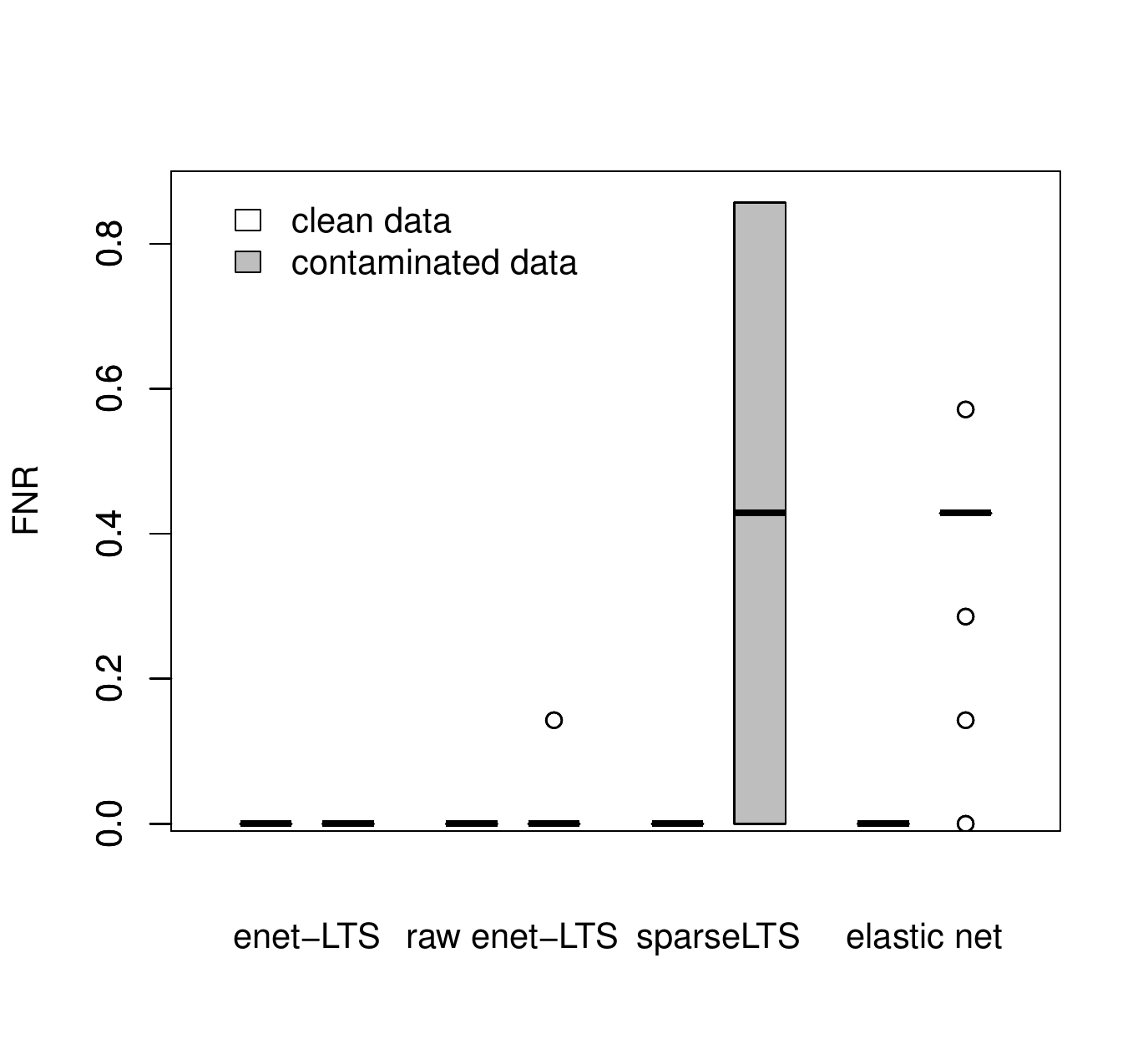}
\hfill
\includegraphics[width=0.5\textwidth]{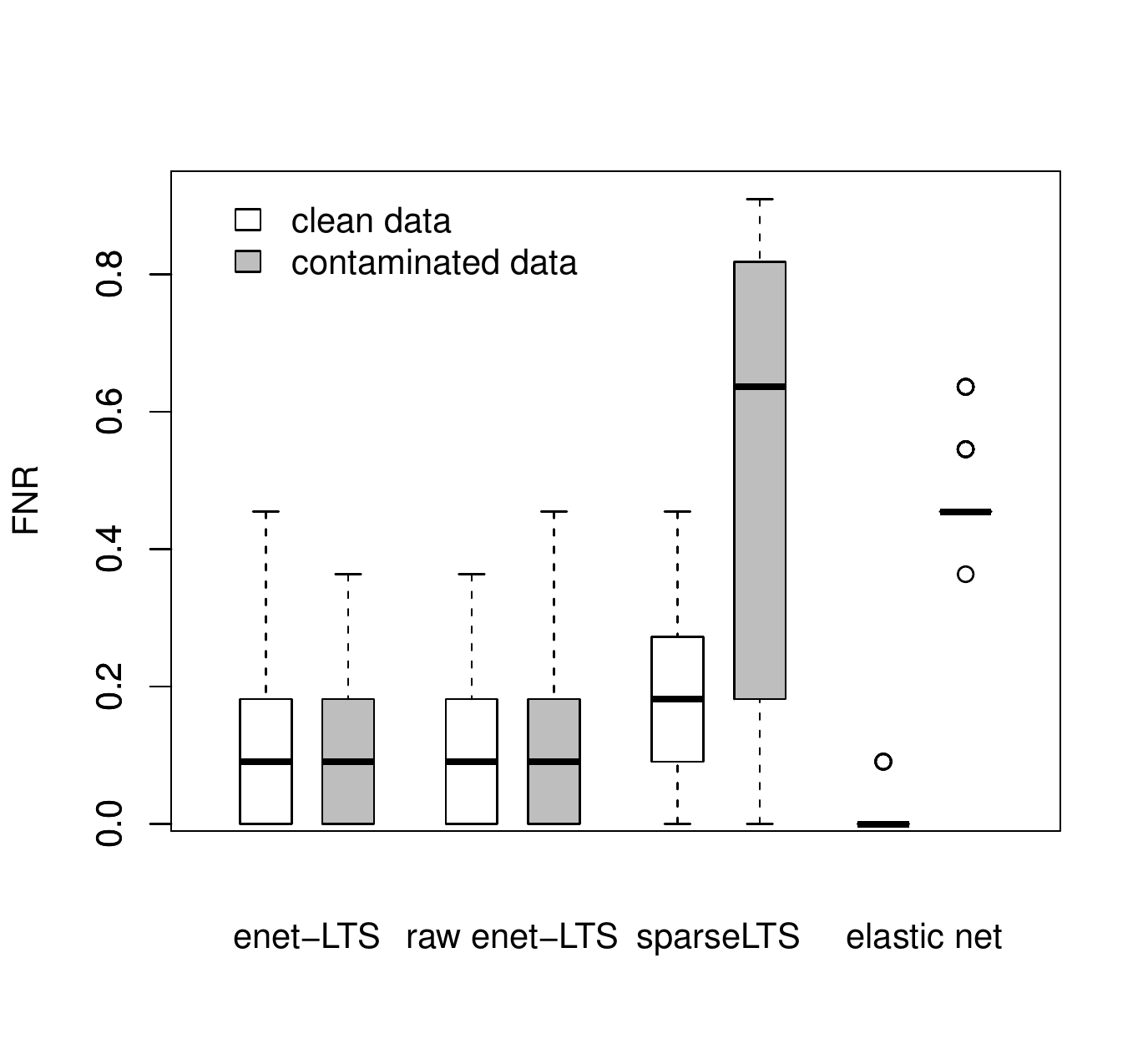}
\caption{False negative rate (FNR) for linear regression.
Left: low dimensional data set ($n=150$ and $p=60$); right: high dimensional data set ($n=50$ and $p=100$).}
\label{fig:fnr_lin}
\end{figure}

\textbf{\textsl{Results for linear regression}:}
The outcome of the simulations for linear regression is summarized in Figures
\ref{fig:rmspe_lin}--\ref{fig:fnr_lin}. The left plots in these figures are for the 
simulations with low dimensional data, and the right plots for the high dimensional
configuration. Figure \ref{fig:rmspe_lin} compares the RMSPE. All methods yield similar
results in the low dimensional non-contaminated case, while in the high dimensional
clean data case the elastic net method is clearly better. However, in the contaminated
case, elastic net leads to poor performance, which is also the case for sparse LTS.
Enet-LTS performs even slightly better with contaminated data, and there is also
a slight improvement visible in the reweighted version of this estimator.
The PRECISION in Figure \ref{fig:bias_lin} shows essentially the 
same behavior. The FPR in Figure \ref{fig:fpr_lin}, reflecting the proportion of incorrectly
added noise variables to the models, shows a very low rate for sparse LTS. Here, the elastic
net even improves in the contaminated setting, and the same is true for enet-LTS.
A quite different picture is shown in Figure \ref{fig:fnr_lin} with the FNR.
Sparse LTS and elastic net miss a high proportion of informative variables in the
contaminated data scenario, which is the reason for their poor overall performance.
Note that the outliers are placed in the informative variables, which seems to be 
particularly difficult for sparse LTS.

\begin{figure}[htbp]
\includegraphics[width=0.5\textwidth]{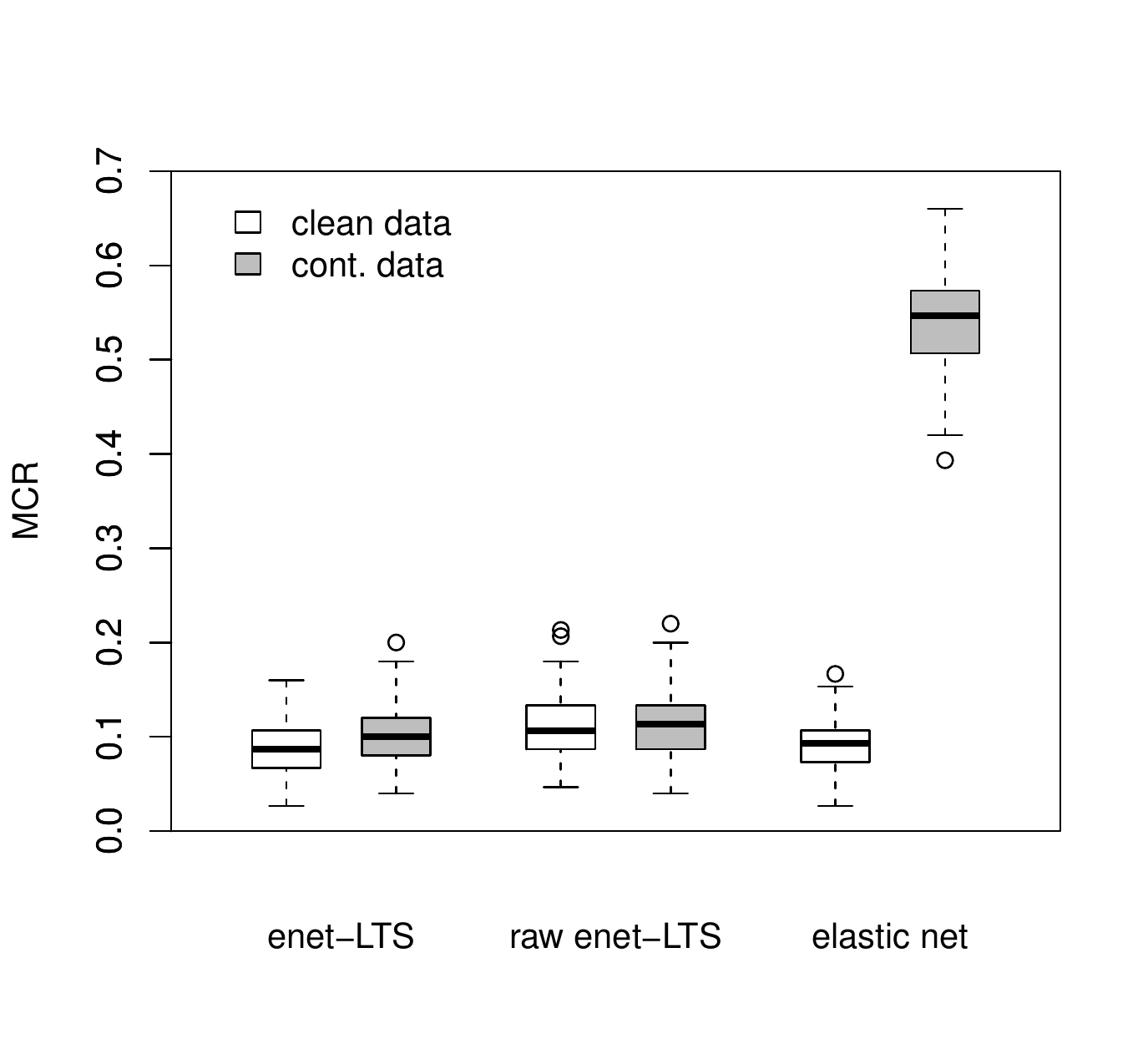}
\hfill
\includegraphics[width=0.5\textwidth]{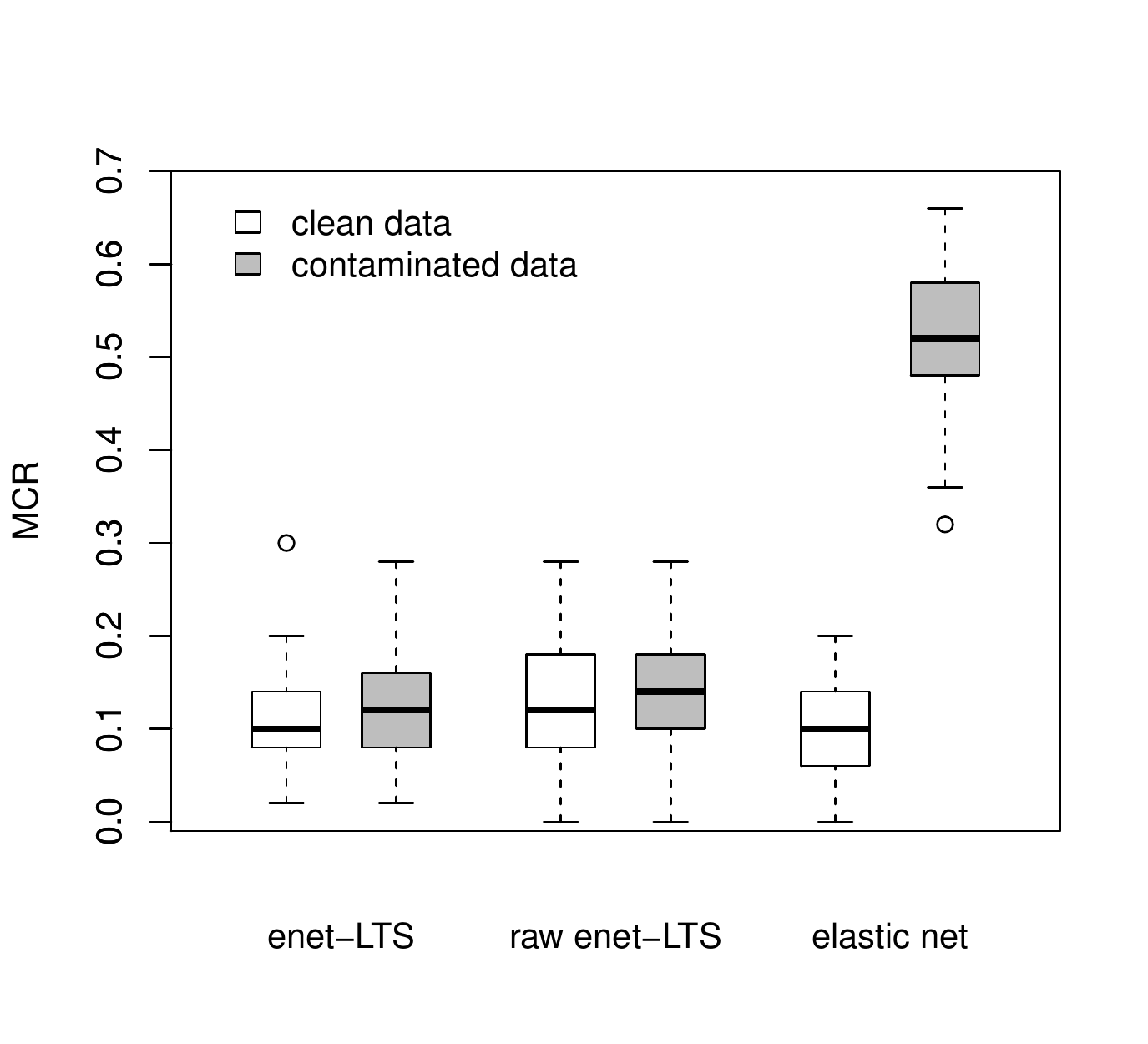}
\caption{Misclassification rate for logistic regression.
Left: low dimensional data set ($n=150$ and $p=50$); right: high dimensional data set ($n=50$ and $p=100$).}
\label{fig:misclas_log}
\end{figure}

\begin{figure}[htbp]
\includegraphics[width=0.5\textwidth]{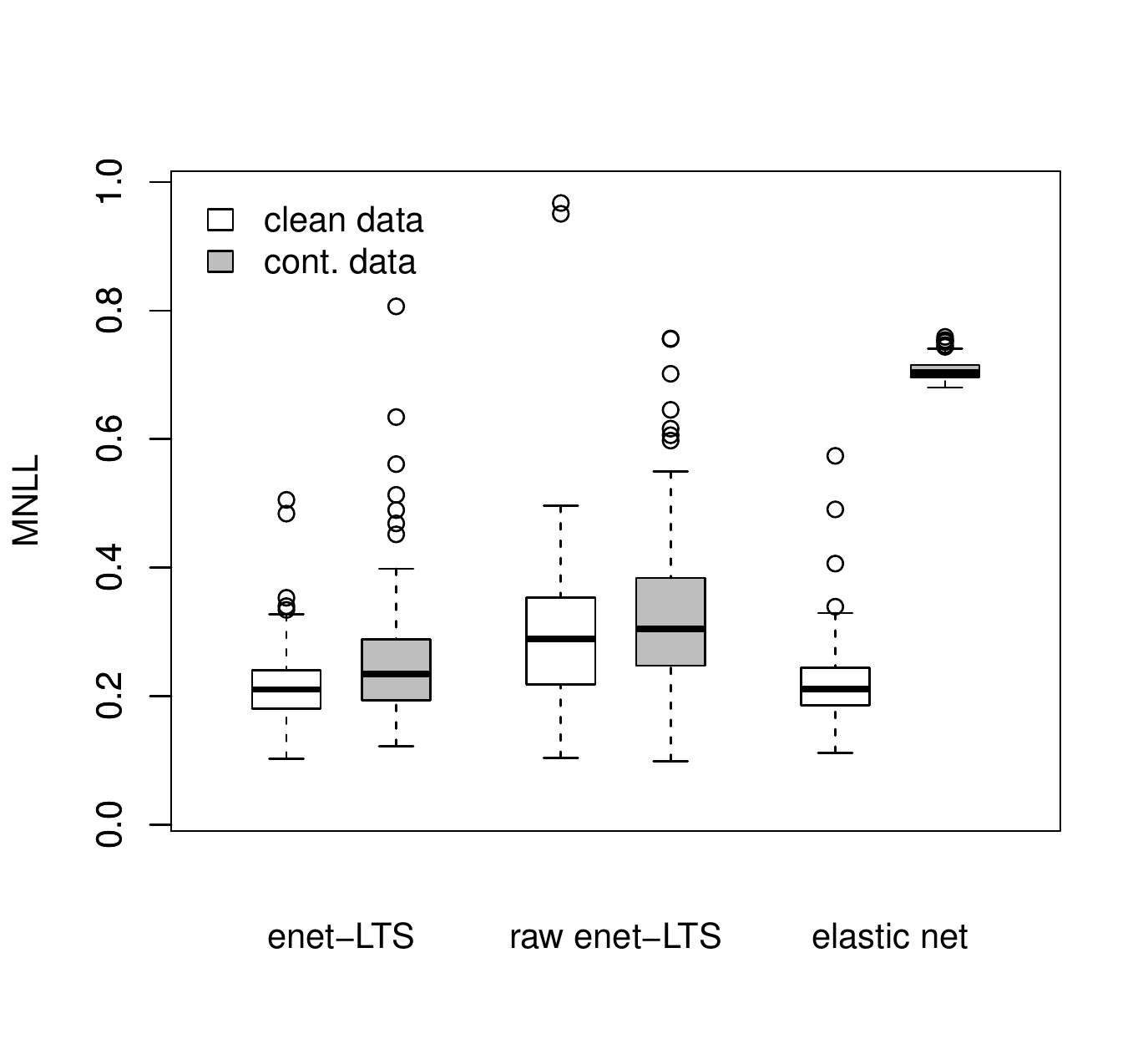}
\hfill
\includegraphics[width=0.5\textwidth]{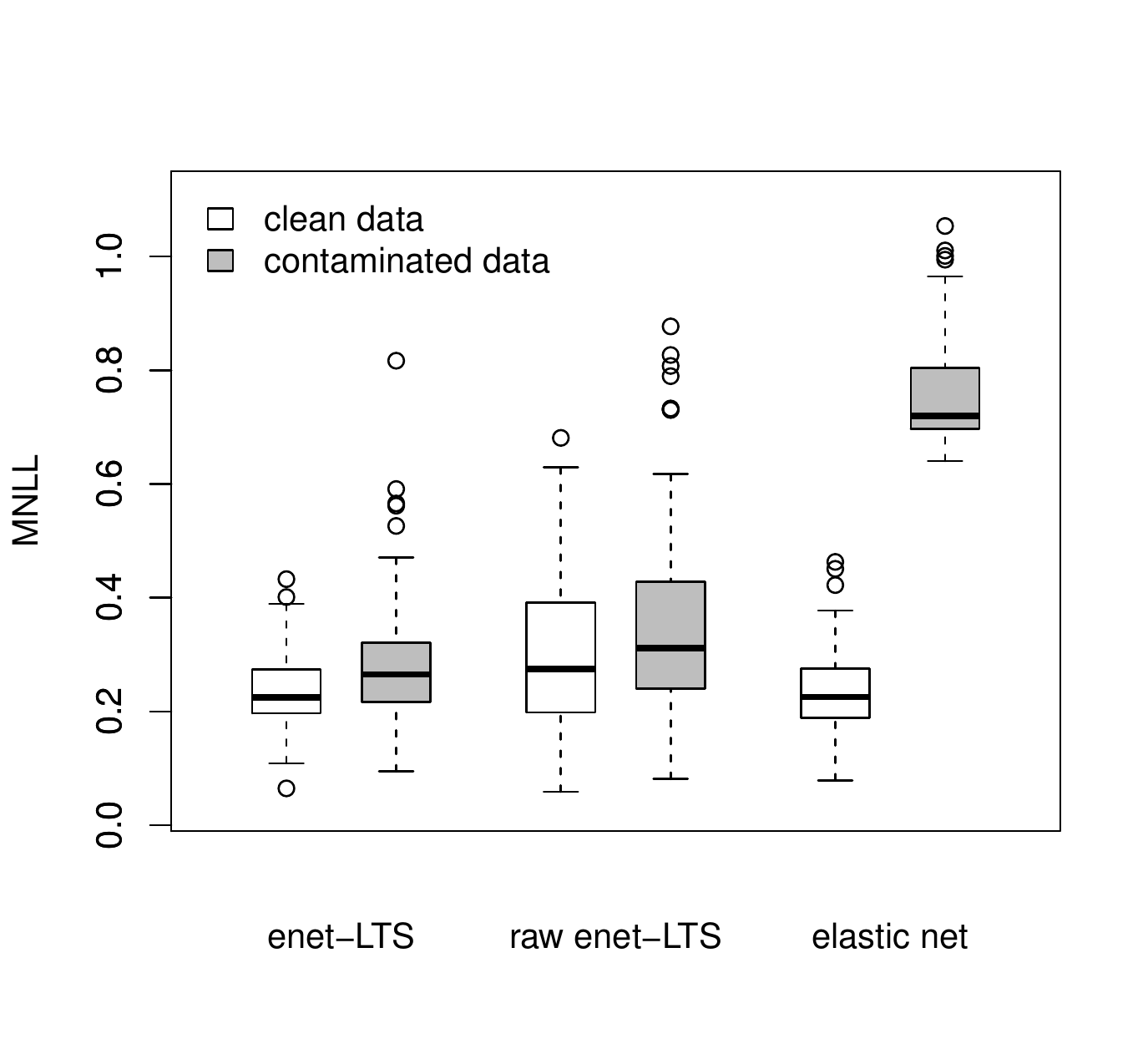}
\caption{The mean of negative likelihood (MNLL) function for logistic regression.
Left: low dimensional data set ($n=150$ and $p=50$); right: high dimensional data set ($n=50$ and $p=100$).}
\label{fig:mnll_log}
\end{figure}

\begin{figure}[htbp]
\includegraphics[width=0.5\textwidth]{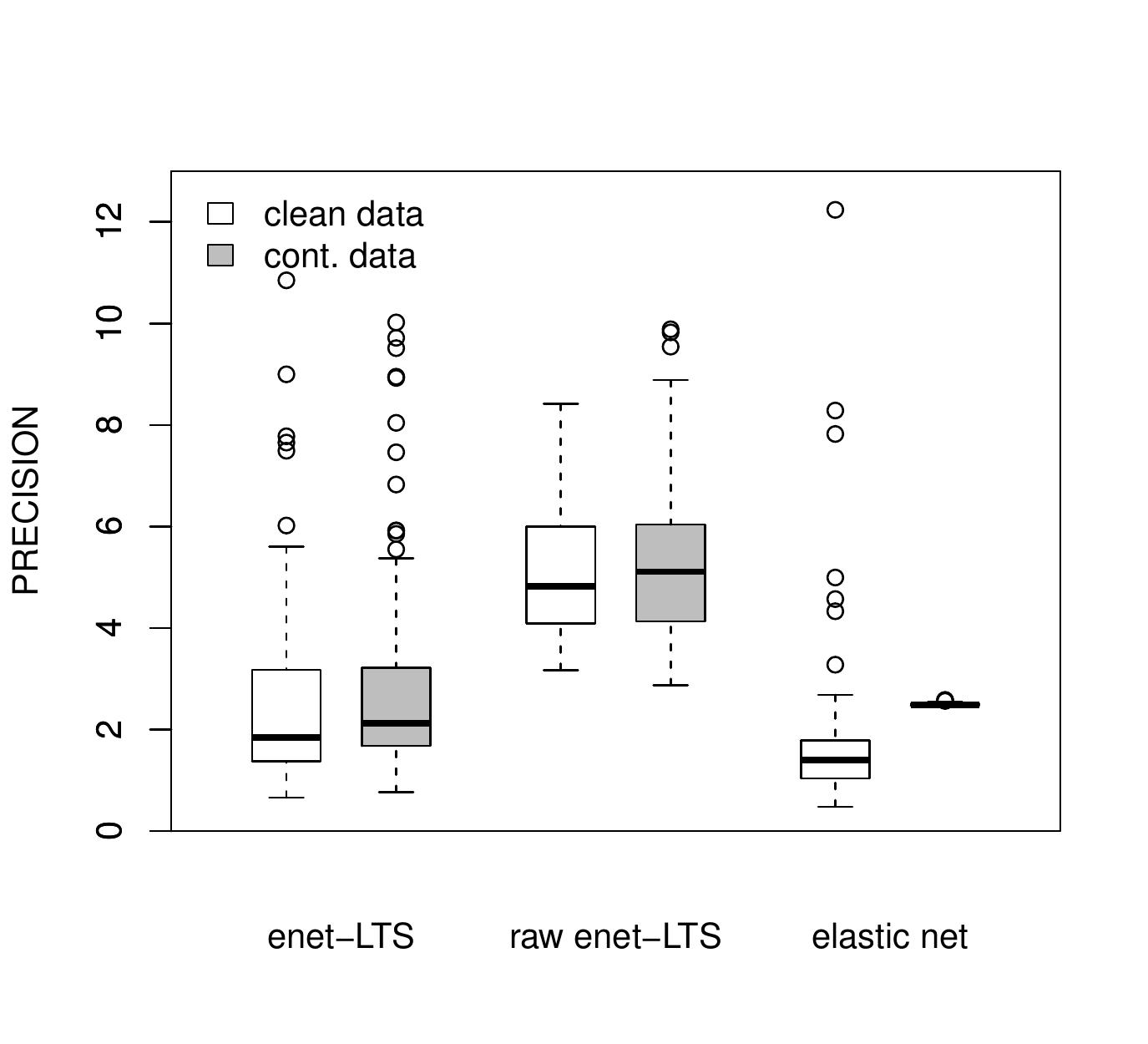}
\hfill
\includegraphics[width=0.5\textwidth]{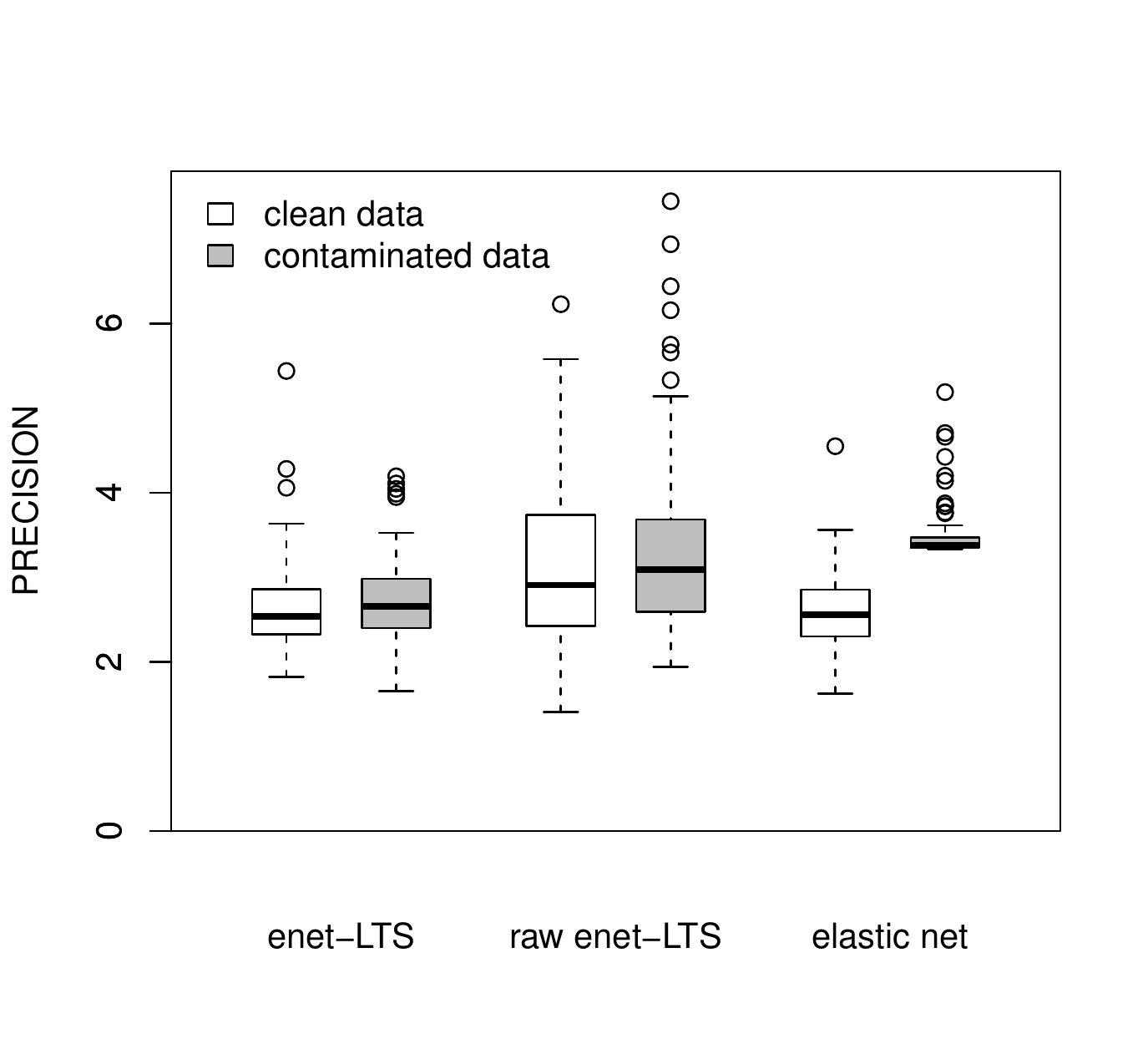}
\caption{Precision of the estimators (PRECISION) for logistic regression.
Left: low dimensional data set ($n=150$ and $p=50$); right: high dimensional data set ($n=50$ and $p=100$).}
\label{fig:bias_log}
\end{figure}

\begin{figure}[htbp]
\includegraphics[width=0.5\textwidth]{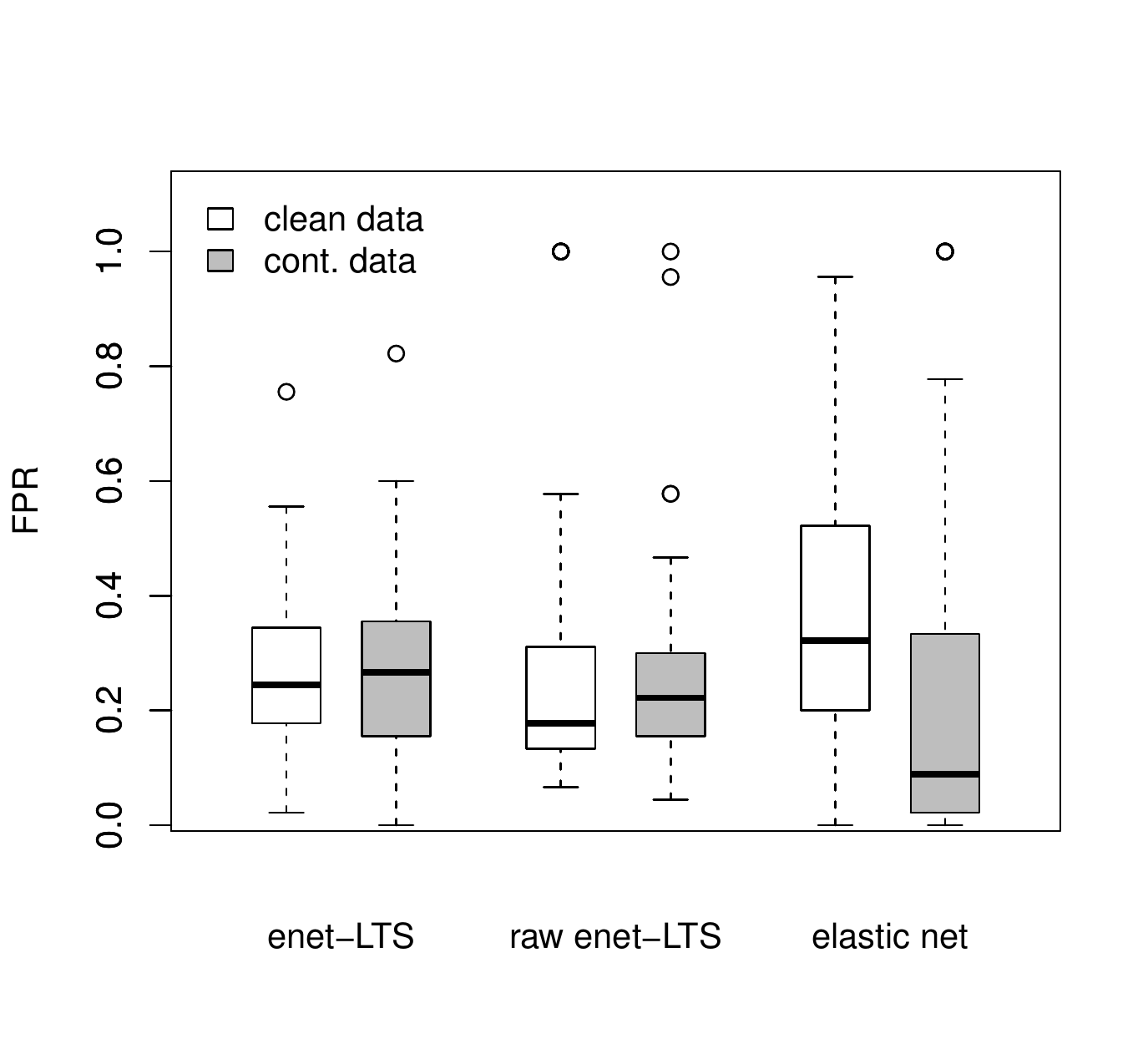}
\hfill
\includegraphics[width=0.5\textwidth]{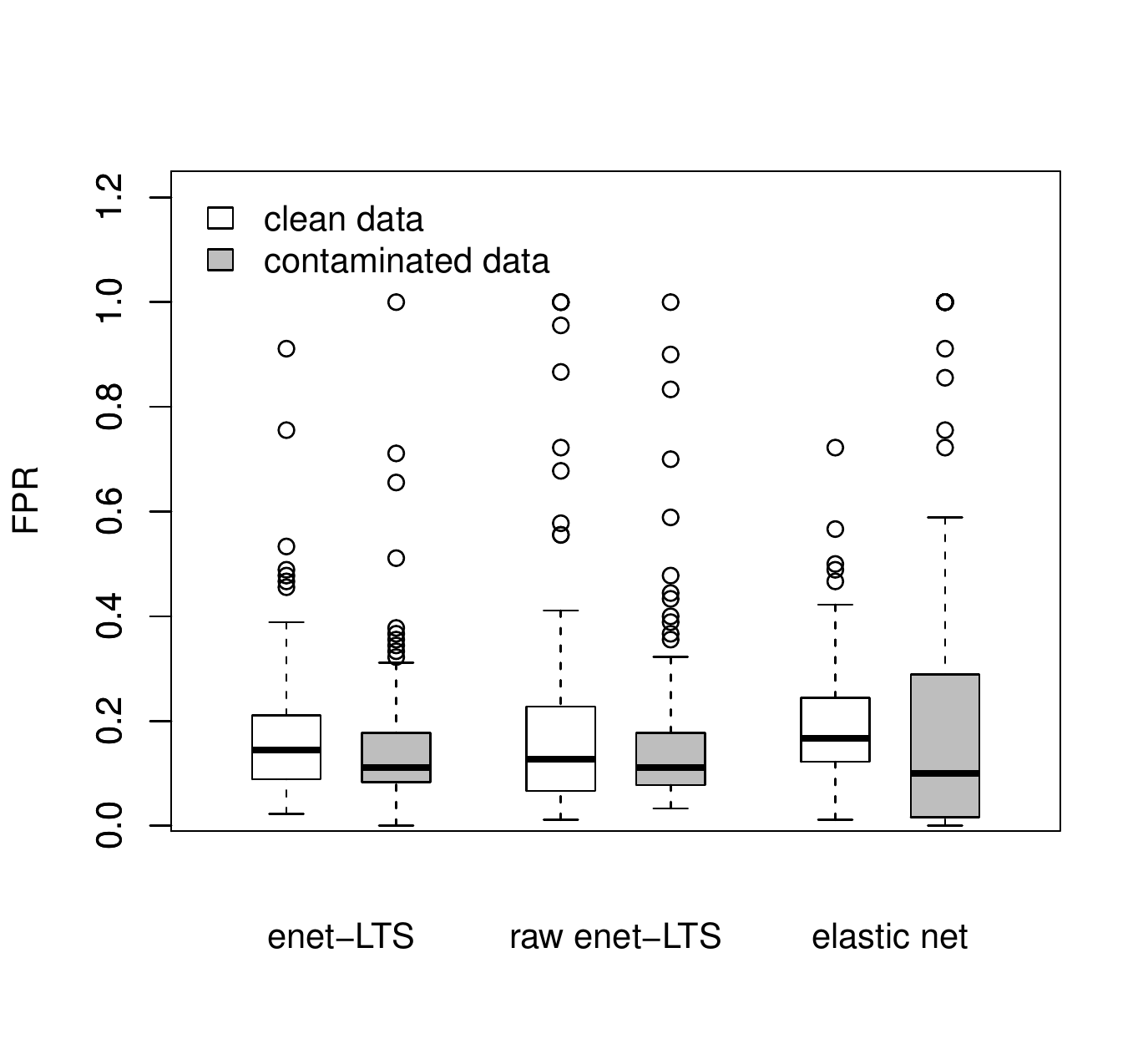}
\caption{False positive rate (FPR) for logistic regression.
Left: low dimensional data set ($n=150$ and $p=50$); right: high dimensional data set ($n=50$ and $p=100$).}
\label{fig:fpr_log}
\end{figure}

\begin{figure}[htbp]
\includegraphics[width=0.5\textwidth]{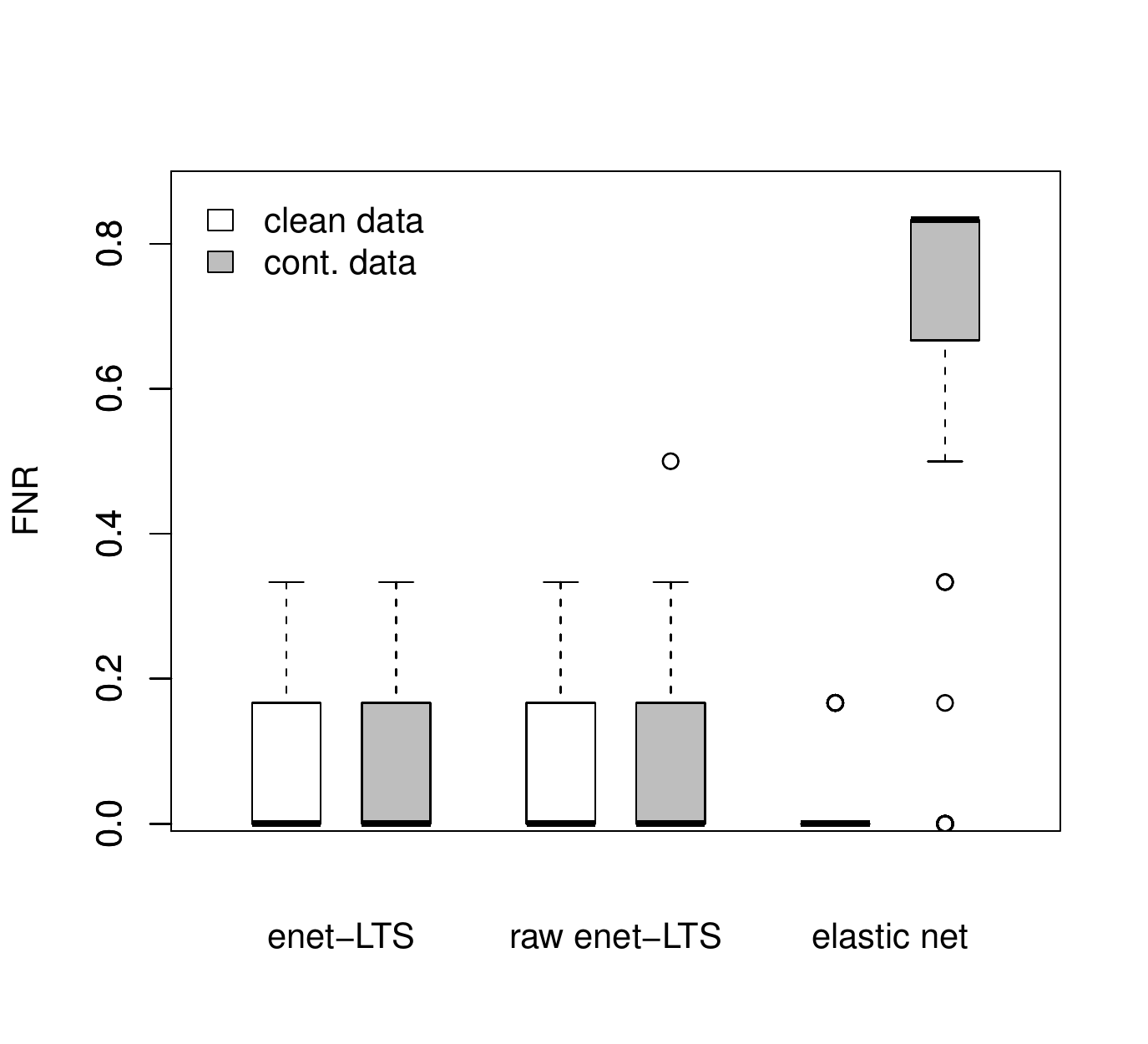}
\hfill
\includegraphics[width=0.5\textwidth]{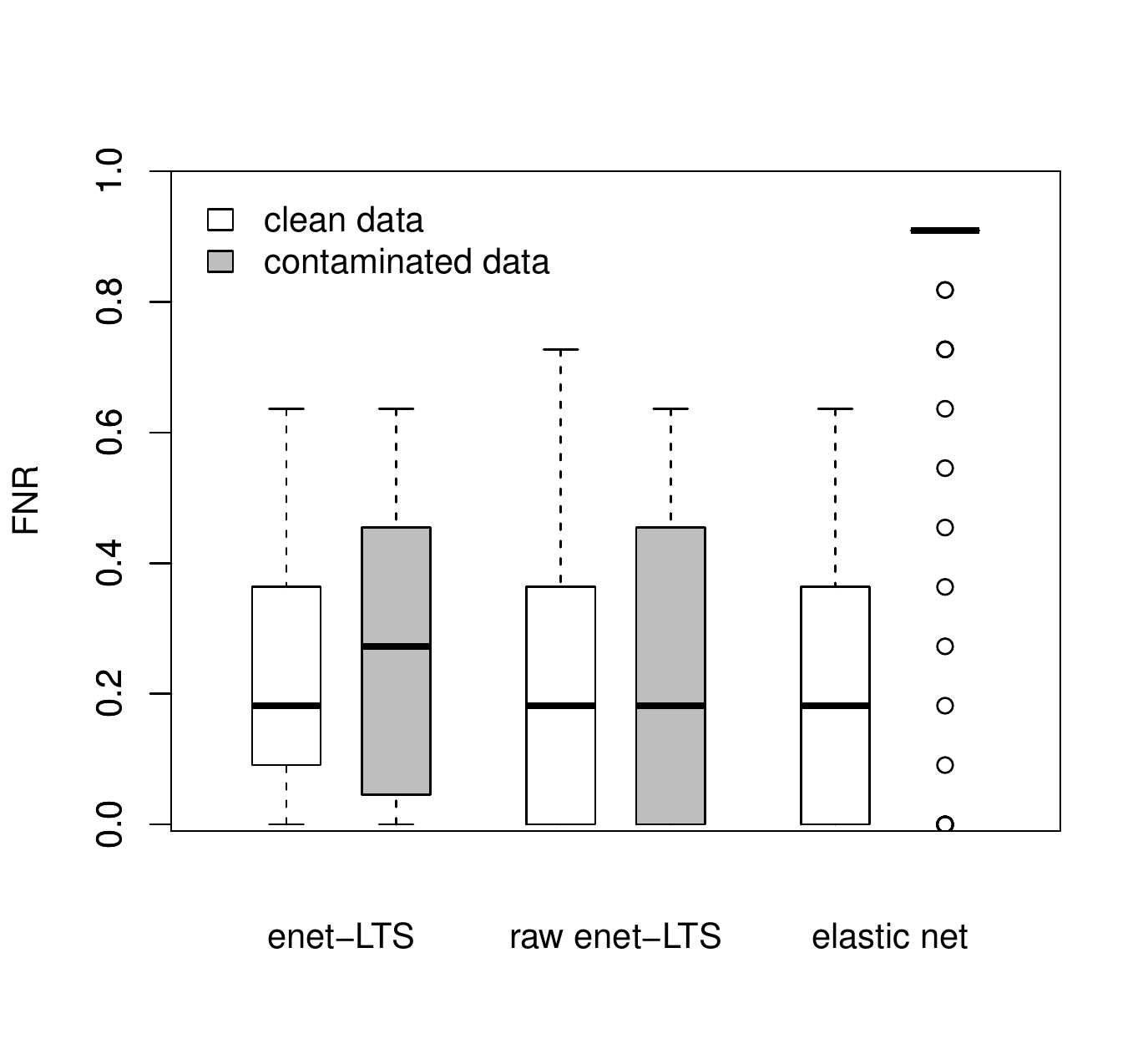}
\caption{False negative rate (FNR) for logistic regression.
Left: low dimensional data set ($n=150$ and $p=50$); right: high dimensional data set ($n=50$ and $p=100$).}
\label{fig:fnr_log}
\end{figure}

\clearpage
\textbf{\textsl{Results for logistic regression}:}
Figures \ref{fig:misclas_log}--\ref{fig:fnr_log} summarize the simulation results for
logistic regression. As before, the left plots refer to the low dimensional case,
and the right plots to the high dimensional data. Within one plot, the results for 
uncontaminated and contaminated data are directly compared. The misclassification rate
in Figure \ref{fig:misclas_log} is around 10\%  for all methods, and it is slightly
higher in the high dimensional situation. 
In case of contamination, however, this rate increases enormously for 
the classical method elastic net.

The average deviances in Figure \ref{fig:mnll_log} show that the reweighting of the 
enet-LTS estimator clearly improves the raw estimate in both the low and high dimensional cases. 
It can also be seen that elastic net is sensitive to the outliers.
The precision of the parameter estimates in Figure \ref{fig:bias_log} reveal a 
remarkable improvement for the reweighted enet-LTS estimator compared to the raw 
version, while there is not any clear effect of the
contamination on the classical elastic net estimator.

The FPR in Figure \ref{fig:fpr_log} shows a certain difference between uncontaminated and
contaminated data for the elastic net, but otherwise the results are quite comparable.
A different picture is visible from the FNR in Figure \ref{fig:fnr_log}, where especially
in the low dimensional case the elastic net is very sensitive to the outliers.
Overall we conclude that the enet-LTS performs very well 
in case of contamination even though this was not clearly visible in the precision, 
and it also yields reasonable results for clean data.

\section{Applications to real data} 
\label{sec:applications}
\vskip-0.25cm 

In this section we focus on applications with logistic regression, and compare the 
non-robust elastic net estimator with the robust enet-LTS method.
The model selection is conducted as described in Section \ref{sec:tuningpara}.
Model evaluation is done with leave-one-out cross validation, i.e.~each observation 
is used as test observation once, a model is estimated on the remaining observations, 
and the negative log-likelihood is calculated for the test observation. 
In these real data examples it is unknown if outliers are present.
In order to avoid an influence of potential outliers on the evaluation of a model,
the 25\% trimmed mean of the negative log-likelihoods is calculated to compare
the models.

\subsection{Analysis of meteorite data}
The time-of-flight secondary iron mass spectroscope COSIMA \cite{kissel2007cosima} 
was sent to the comet Churyumov-Gerasimenko in the Rosetta space mission by the ESA 
to analyze the elemental composition of comet particles which were collected there 
\cite{schulz2015comet}. As 
reference measurements, samples of meteorites provided by the Natural History Museum 
Vienna were analyzed with the same type of spectroscope at Max Planck Institute for Solar 
System Research in G\"ottingen.

Here we apply our proposed method for logistic regression to the measurements from particles 
from the meteorites Ochansk and Renazzo with 160 and 110 spectra, respectively. We restrict 
the mass range to 1-100mu, consider only mass windows where inorganic and organic ions
can be expected as described in \cite{Varmuza11} and variables with positive median absolute deviation.
So we obtain $p=1540$ variables. Further, the data is normalized to have constant row sum 100.

Table \ref{tab:meteorite} summarizes the results for the comparison of the methods. 
The trimmed MNLL is much smaller for the enet-LTS estimator than for the classical elastic 
net method. The reweighting step improves the quality of the model further. The selected 
tuning parameter $\alpha_{opt}$ is much smaller for enet-LTS than for the classical elastic 
net method which strongly influences the number of variables in the models.
\begin{table}[htb]
\centering
\begin{tabular}{lrr}
  \hline
 & number variables & trimmed MNLL \\ 
  \hline
elastic net & 136 & 0.00866 \\
  enet-LTS raw & 294 & 0.00030 \\ 
  enet-LTS & 397 & 0.00014 \\ 
   \hline
\end{tabular}
\caption{Renazzo and Ochansk: Number of variables in the optimal models and trimmed mean negative log-likelihood from leave-one-out cross validation of the optimal models.}
\label{tab:meteorite}
\end{table}

\begin{figure}[htbp]
\centering
\includegraphics[width=0.6\textwidth]{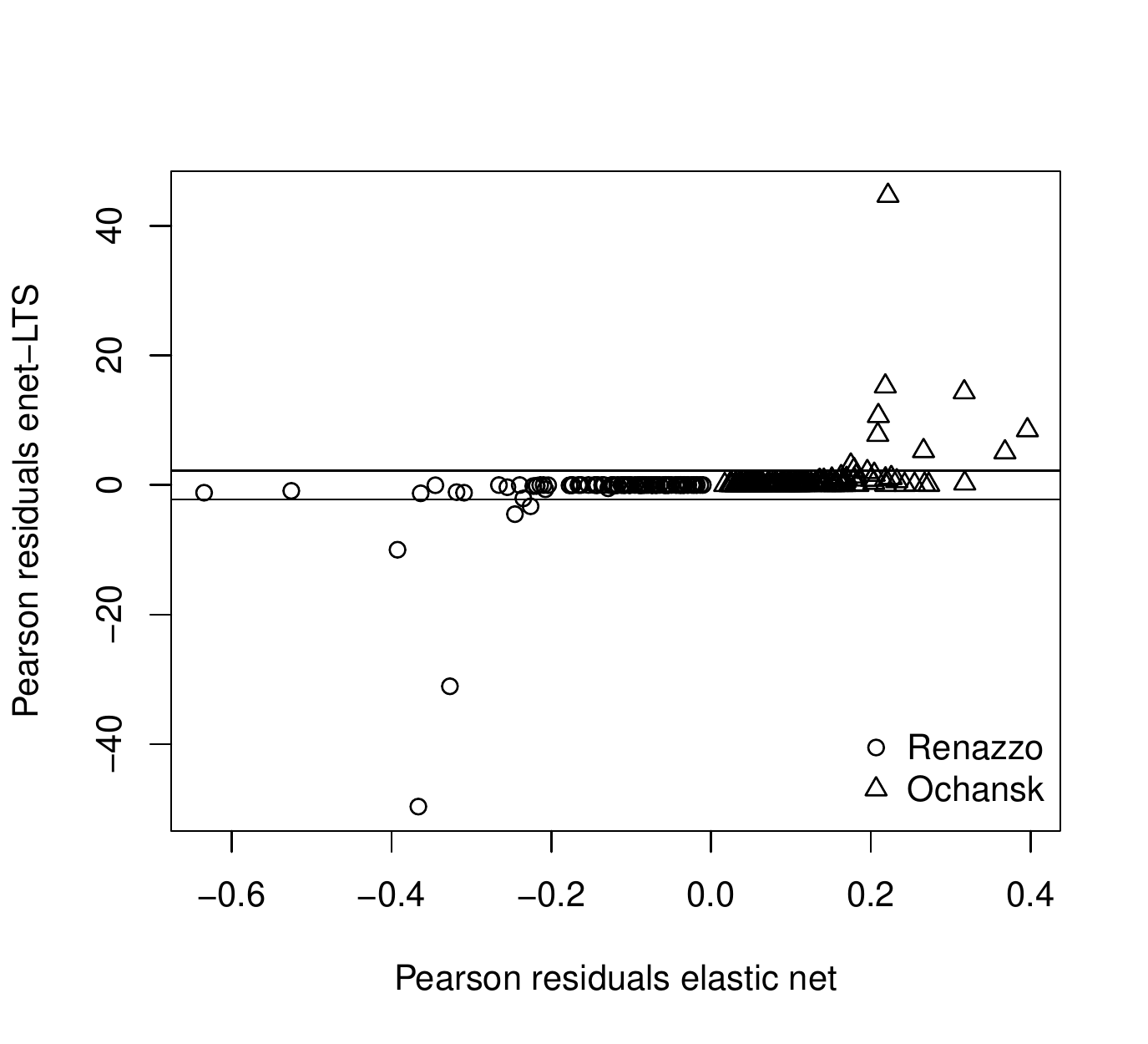}
\caption{Renazzo and Ochansk: the Pearson residuals of elastic net and the raw enet-LTS estimator. The horizontal lines indicate the 0.0125 and the 0.9875 quantiles of the standard normal distribution. }
\label{fig:meteorite}
\end{figure}

\begin{figure}[htbp]
  \centering
\includegraphics[width=\textwidth]{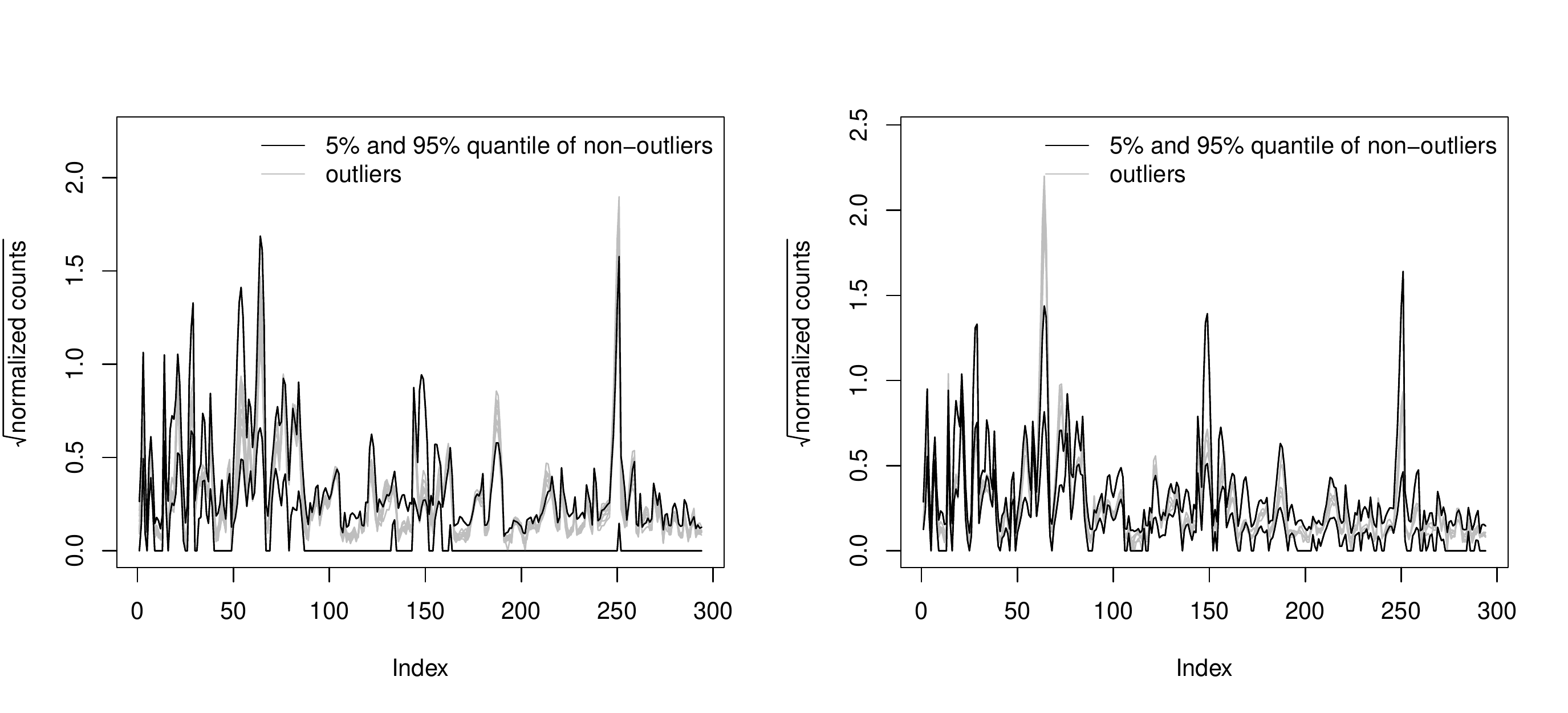}
\caption{The index refers to the index of the variables included in 
the model of raw enet-LTS. The detected outliers are visualized by grey lines, 
while the black lines represent the 5\% and 95\% quantile of the non-outlying spectra for 
Ochansk (left) and Renazzo (right).}\label{fig:meteorite2}
\end{figure}

Figure \ref{fig:meteorite} compares the Pearson residuals of the elastic net model and 
the enet-LTS model. In the classical approach no abnormal observations can be detected. 
With the enet-LTS model several observations are identified as outliers by the 1.25\% and 98.25\% quantiles of the standard normal distribution, which are marked as horizontal lines in Figure \ref{fig:meteorite}. Closer investigation 
showed that these spectra lie on the outer border of the measurement area and are potentially 
measured on the target instead of the meteorite particle. Their multivariate structure for those 
variables
which are included in the model is visualized in Figure \ref{fig:meteorite2}, where we can see 
that in some variables they have particularly large values compared to the majority of the group.

\subsection{Analysis of the glass vessels data}

Archaeological glass vessels where analyzed with electron-probe X-ray micro-analysis to 
investigate the chemical concentrations of elements in order to learn more about their 
origin and the trade market at the time of their making in the 16$^{th}$ and 17$^{th}$ century \cite{Janssens98}.
Four different groups were identified, i.e. sodic, potassic, 
potasso-calcic and calcic glass vessels.
For demonstration of the performance of logistic regression, two groups are selected 
from the glass vessels data set. The first group is the potassic group with 15 spectra, 
the second group the potasso-calcic group with 10 spectra. As in \cite{Filzmoser08} 
we remove variables with MAD equal to zero, resulting in $p=1905$ variables.

The quality of the selected models is described in Table \ref{tab:glassvessel}. 
The trimmed mean of the negative log likelihoods is much smaller for enet-LTS than 
for elastic net. The reweighting step in enet-LTS hardly improves the model, 
but includes more variables. Again, both enet-LTS models include more variables 
than the elastic net model. In the elastic net model the penalty gives higher emphasis 
on the $l_1$ term, i.e. $\alpha_{opt}=0.8$; for enet-LTS it is $\alpha_{opt}=0.05$. 
\begin{table}[htb]
\centering
\begin{tabular}{lrr}
  \hline
 & number variables & trimmed MNLL \\ 
  \hline
elastic net & 50 & 0.004290 \\ 
  enet-LTS raw & 375 & 0.000345 \\ 
  enet-LTS & 448 & 0.000338 \\ 
   \hline
\end{tabular}
\caption{Glass vessel data: number of variables in the optimal models, 
and trimmed mean negative log-likelihood from leave-one-out cross validation of the optimal models.}
\label{tab:glassvessel}
\end{table}

\begin{figure}[htbp]
\includegraphics[width=\textwidth]{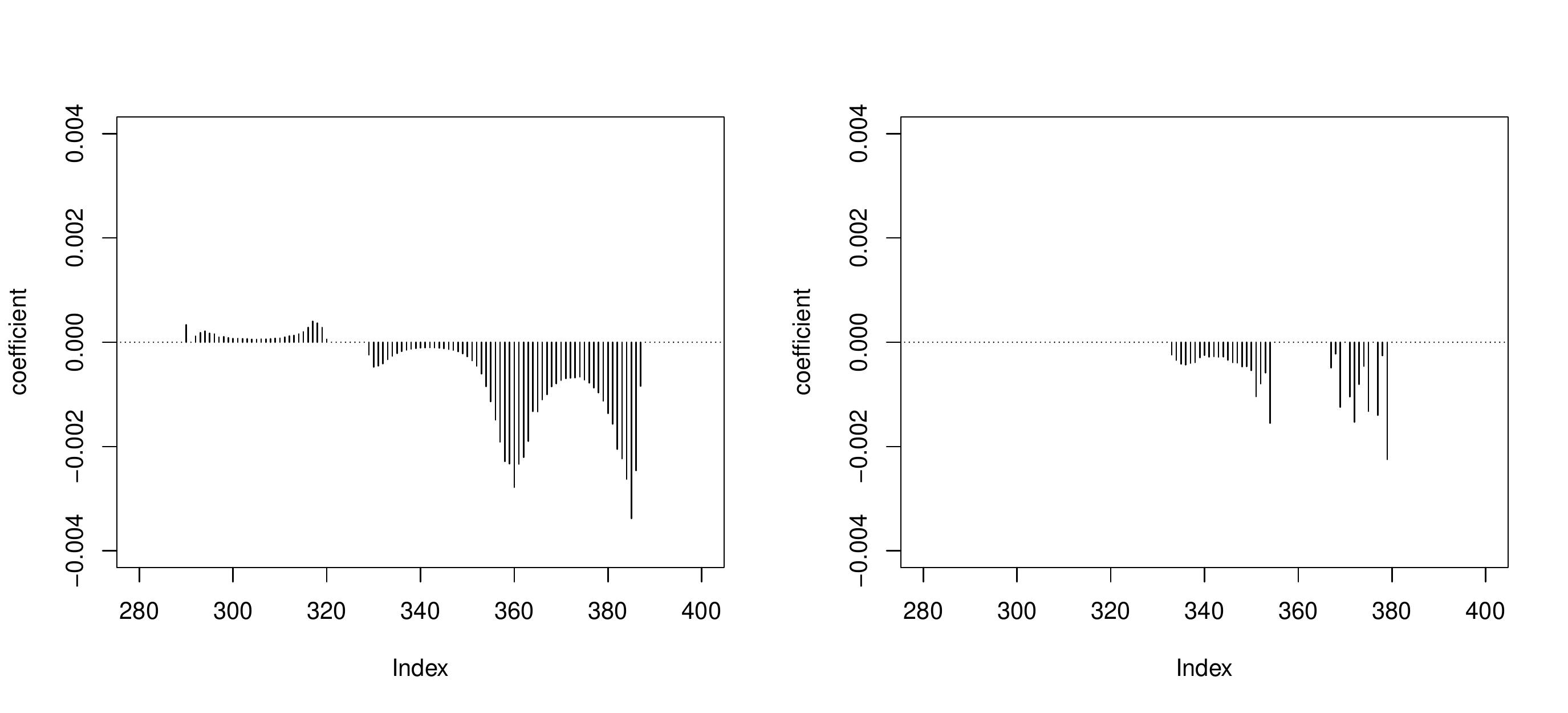}
\caption{Glass vessels: coefficient estimate of the reweighted enet-LTS model 
(left) and 
coefficient estimate of the elastic net mode (right) for a selected variable range.}\label{fig:glassvessel}
\end{figure}

Different behavior of the coefficient estimates can be expected. 
Figure~\ref{fig:glassvessel} (left) shows coefficients of the reweighted enet-LTS 
model corresponding to variables associated with potassium and calcium.  
The band which is associated with potassium has positive coefficients, i.e. high 
values of these variables correspond to the potassic group which is coded with 
ones in the response. High values of the variables in the band which is associated 
with calcium will favor a classification to the potasso-calcic group (coded with zero), 
since the coefficients for these variables are negative. Further, 
it can be observed that neighboring variables, which are correlated, have similar 
coefficients. This is favored by the $l_2$ term in the elastic net penalty. 
In Figure~\ref{fig:glassvessel} (right) the coefficient estimates of the elastic net model 
are visualized. Fewer coefficients are non-zero than for enet-LTS which was favored 
by the $l_1$ term in the elastic net penalty, but in the second block of non-zero coefficients 
neighboring variables receive very different coefficient estimates. 

\section{Computation time}
\label{sec:comptime}
\vskip-0.25cm 

\begin{figure}[htbp]
\includegraphics[width=0.5\textwidth]{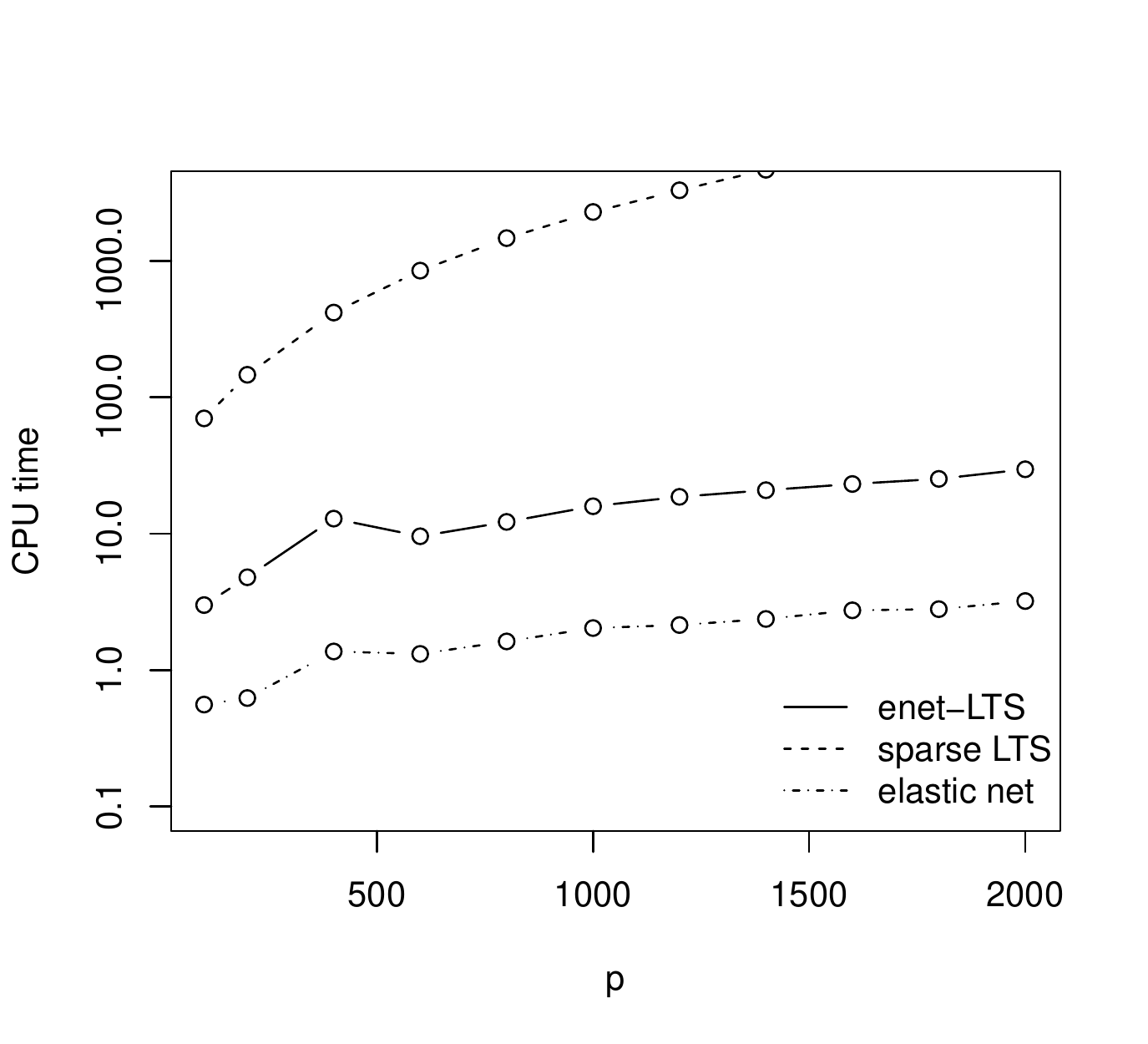}
\hfill
\includegraphics[width=0.5\textwidth]{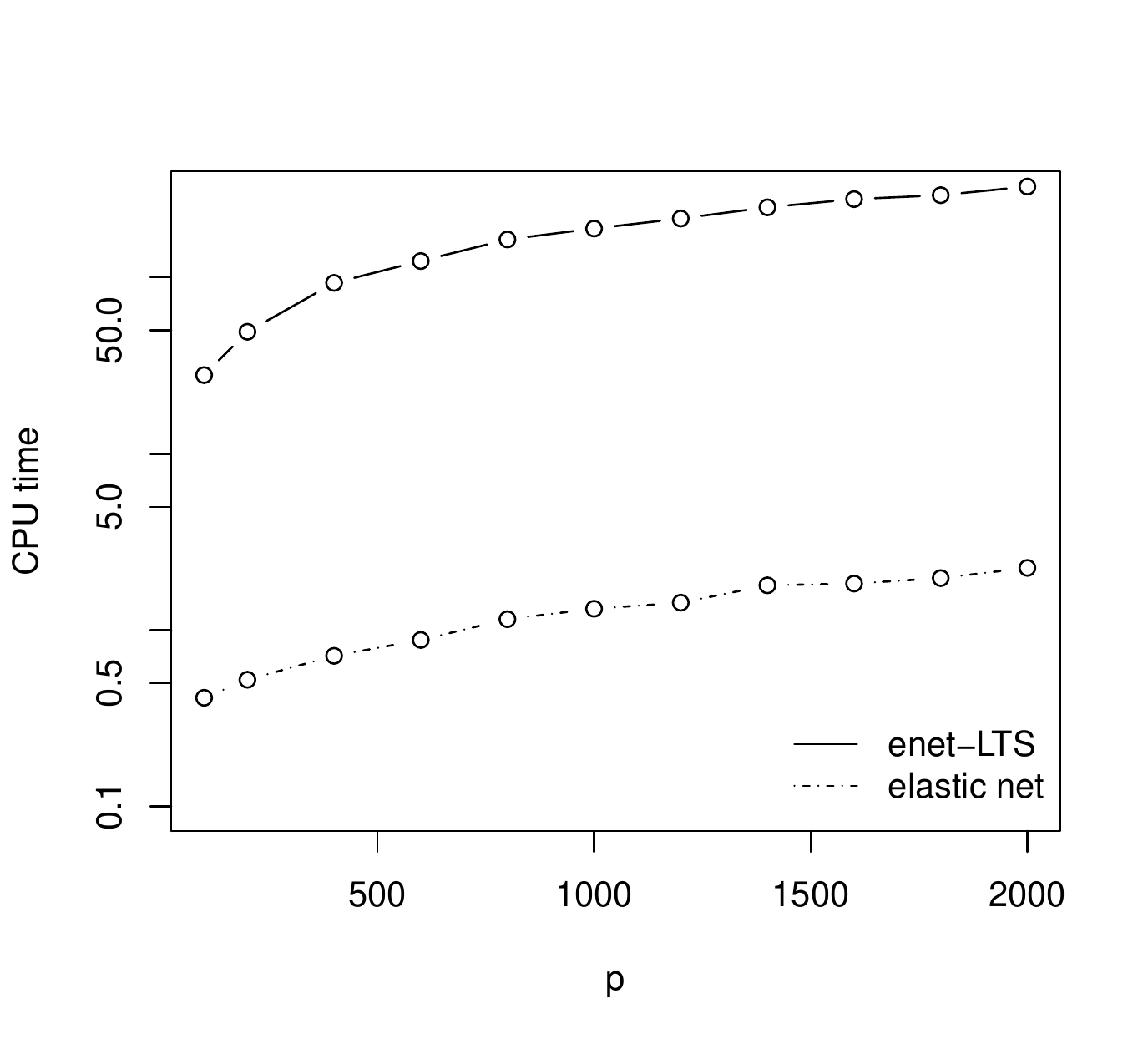}
\caption{CPU time in seconds (log-scale), averaged over 5 replications, for fixed $n=150$ and 
varying $p$; left: for linear regression; right: for logistic regression.}
\label{fig:compt_time}
\end{figure}

\begin{figure}[htbp]
\includegraphics[width=0.5\textwidth]{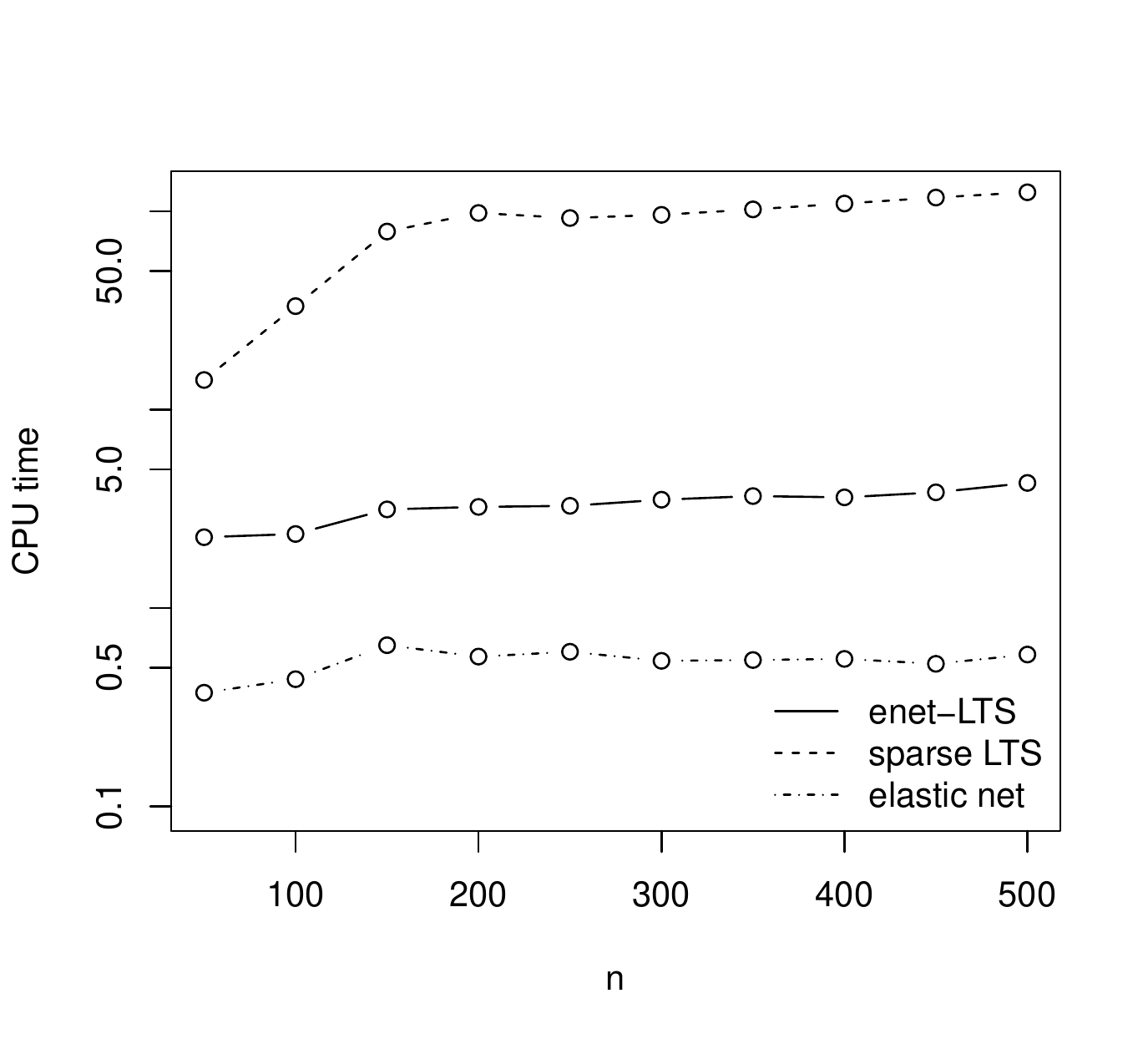}
\hfill
\includegraphics[width=0.5\textwidth]{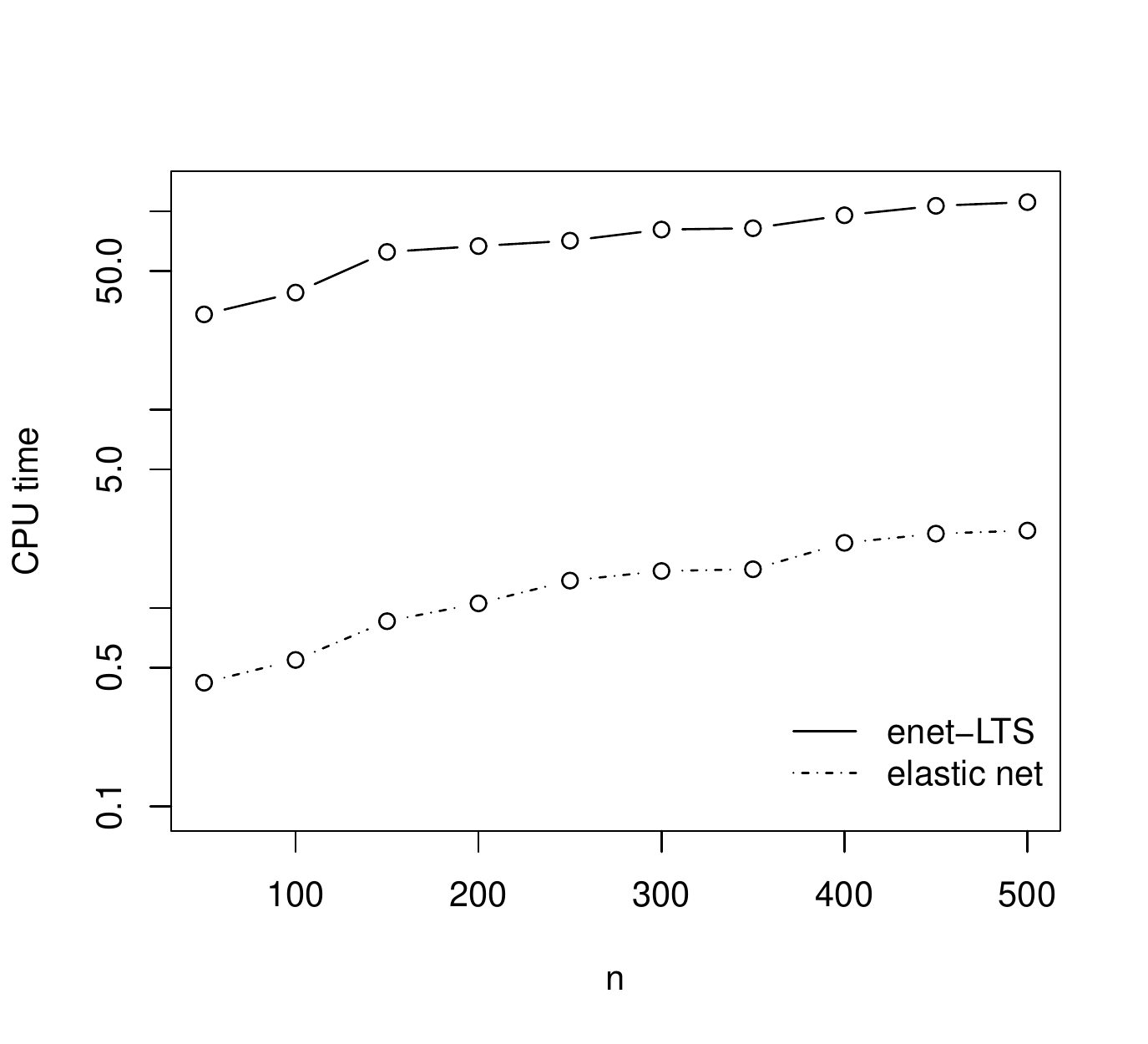}
\caption{CPU time in seconds (log-scale), averaged over 5 replications, for fixed $p=100$ and 
varying $n$; left: for linear regression; right: for logistic regression.}
\label{fig:compt_time_n}
\end{figure}

For our algorithm we employ the classical elastic net estimator as it is 
implemented in the R package $glmnet$ \cite{FriedmanR16}. So, it is natural to
compare the computation time of our algorithm 
with this method. In the linear regression case we also compare with
the sparse LTS estimator implemented in the R package $robustHD$ \cite{AlfonsR13}.
For calculating the estimators we take a grid of five values for both tuning parameters 
$\alpha$ and $\lambda$. The data sets are 
simulated as in Section \ref{sec:simulations} for a fixed number of observations $n=150$,
but for a varying number of variables $p$ in a range from $50$ to $2000$.
In Figure~\ref{fig:compt_time} (left: linear regression, right: logistic regression), 
the CPU time is reported in seconds, as an average over 
$5$ replications. In order to show the dependency on the number of 
observations $n$, we also 
simulated data sets for a fixed number of variables $p=100$ with a varying number of observations $n=50,100,\dots,500$. The results for linear and logistic regression are summarized 
in Figure~\ref{fig:compt_time_n}. The computations 
have been performed on an Intel Core 2 Q9650 @ $3000$ GHz$\times$4 processor.

Let us first consider the dependency of the computation time on the number of 
variables $p$ for linear regression, shown in the left 
plot of Figure~\ref{fig:compt_time}. 
Sparse LTS increases strongly with the number of variables $p$ since it is based on the LARS 
algorithm which has a computational complexity of $\mathcal{O}(p^3+np^2)$ \cite{Efron2004}.
Also for the smallest
number of considered variables, the computation time is considerably higher than for the other
two methods. The reason is that for each value of $\lambda$ and each step in the CV the best subset is determined starting
 with 500 elemental subsets. In this setting at least 25,000 
estimations of a Lasso model are needed, because for each cross validation step at each 
of the 5 values of $\lambda$,
two C-steps for 500 elemental subsets are carried out, 
and for the 10 subsamples with lowest value of the objective function, further C-steps
are performed.
In contrast, the enet-LTS estimator starts with 500 elemental subsets only for one 
combination of $\alpha$ and $\lambda$,
and takes the \textit{warm start} strategy for subsequent combinations. This saves 
computation time, and there is indeed only a slight increase with $p$ visible when compared to
the elastic net estimator. In total approximately 1,700 
elastic net models are estimated in this procedure, which are considerably fewer than for the sparse LTS approach.
The computation time of sparse LTS also increases with $n$ due to the computational
complexity of LARS, while the increase is only minor for enet-LTS, see 
Figure~\ref{fig:compt_time_n} (left).

The results for the computation time in logistic regression are presented in 
Figure \ref{fig:compt_time} (right) and \ref{fig:compt_time_n} (right). 
Here we can only compare the classical elastic net estimator and the proposed 
robustified enet-LTS version. The difference in computation time between elastic net 
and enet-LTS is again due to the many calls of the {\tt glmnet} function within enet-LTS. 
The robust estimator is considerably slower in logistic regression when compared to linear 
regression for the same number of explanatory variables or observations. The reason is
that more C-steps are necessary to identify the optimal subset for each parameter 
combination of $\alpha$ and $\lambda$.

\section{Conclusions} 
\label{sec:conclude}
\vskip-0.25cm 

In this paper, robust methods for linear and logistic regression using the elastic net penalty were
introduced. This penalty allows for variable selection, can deal with high
multicollinearity among the variables, and is thus very appropriate in high dimensional
sparse settings. Robustness has been achieved by using trimming. This usually leads to
a loss in efficiency, and therefore a reweighting step was introduced. Overall, the 
outlined algorithms for linear and logistic regression turned out to yield good performance
in different simulation settings, but also with respect to computation time.
Particularly, it was shown that the idea of using ``warm starts'' for parameter
tuning allows to save computation time, while the precision is still preserved.
A competing method for robust high dimensional linear regression, the sparse LTS estimator 
\cite{AlfonsR13}, does not use this idea, and is thus much less attractive concerning
computation time, especially in case of many explanatory variables. We should also
admit that for other simulation settings (not shown here), the difference
between sparse LTS and the enet-LTS estimator is not so big, or even marginal, depending
on the exact setting.

For this reason, a further focus was on the robust high dimensional logistic regression
case. 
We consider such a method as highly relevant, since in many modern applications in
chemometrics or bio-informatics, one is confronted with data information from two groups,
with the task to find a classification rule and to identify marker variables which support
the rules. Outliers in the data are frequently a problem, and they can affect the 
identification of the marker variables as well as the performance of the classifier.
For this reason it is desirable to treat outliers appropriately. It was shown in simulation
studies as well as in data examples, that in presence of outliers the new proposal
still works well, while its classical non-robust counterpart can lead to poor 
performance.

Note that in \cite{Park16} a logistic regression method with elastic net penalty is proposed using 
weights to reduce the influence of outliers. Their approach is to perform outlier 
detection in a PCA space, obtain weights based on robust Mahalanobis distances in the PCA 
score space and derive weights from these distances. These weights are then used to 
down-weight the negative log likelihoods in the penalized objective function to reduce 
the influence of outliers. However, it is not guaranteed that outliers can be detected 
in the PCA score space. An increasing number of uninformative variables will disguise 
observations deviating from the majority only in few informative variables, but these 
hidden outlying observations can still distort the model. Therefore, model based outlier 
detection is highly recommended as proposed in our algorithm.

The algorithms to compute the proposed estimators are implemented in R functions,
which are available upon request from the authors.
The basis for the computation of the robust estimator is the R package $glmnet$ 
\cite{FriedmanR16}. This package also implements the case of multinomial and 
Poisson regression. Naturally, a further extension of the algorithms introduced
here could go into these directions.
Further work will be devoted to the theoretical properties of the family of enet-LTS 
estimators.

\section*{Acknowledgments}

This work is partly supported by the Austrian Science Fund (FWF), project P 26871-N20 and 
by grant TUBITAK 2214/A from the Scientific and Technological Research Council
of Turkey (TUBITAK). 

The authors thank F. Brandst\"atter, L. Ferri\`ere, and C. Koeberl (Natural History Museum
Vienna, Austria) for providing meteorite samples, C. Engrand (Centre de Sciences 
Nucl\'eaires
et de Sciences de la Mati\`ere, Orsay, France) for sample preparation, and M. Hilchenbach
(Max Planck Institute for Solar System Research, G\"ottingen, Germany) for TOF-SIMS
measurements.
The authors are grateful to Kurt Varmuza for valuable feedback on the results of
the meteorite data.


\end{document}